\documentclass[%
11pt,
onecolumn,preprintnumbers,pre,
 nofootinbib,
 amsmath,amssymb,
]{revtex4-1}

\usepackage[english]{babel}
\usepackage[utf8]{inputenc}
\usepackage[babel]{csquotes}
\usepackage[T1]{fontenc}
\usepackage{amsthm}
\usepackage{amsmath}
\usepackage{amsfonts}
\usepackage{mathrsfs}
\usepackage{amssymb}
\usepackage{mathtools}
\usepackage{graphicx}
\usepackage{xcolor}
\usepackage{subfigure}
\usepackage{geometry}
\usepackage{braket}
\usepackage{sidecap}
\usepackage{float}
\usepackage{bm}
\usepackage{lipsum}

\theoremstyle{remark}

\begin{document}
\title{Generating directed networks with prescribed Laplacian spectra}

\author{Sara Nicoletti}
\affiliation{Dipartimento di Ingegneria dell'Informazione, Universit\`{a} di Firenze,
Via S. Marta 3, 50139 Florence, Italy}
\affiliation{Universit\`{a} degli Studi di Firenze, Dipartimento di Fisica e Astronomia,
CSDC and INFN, via G. Sansone 1, 50019 Sesto Fiorentino, Italy}
\author{Timoteo Carletti}
\affiliation{naXys, Namur Institute for Complex Systems, University of Namur, 8 Rempart de la Vierge, B5000 Namur, Belgium}
\author{Duccio Fanelli}
\affiliation{Universit\`{a} degli Studi di Firenze, Dipartimento di Fisica e Astronomia,
CSDC and INFN, via G. Sansone 1, 50019 Sesto Fiorentino, Italy}
\author{Giorgio Battistelli}
\affiliation{Dipartimento di Ingegneria dell'Informazione, Universit\`{a} di Firenze,
Via S. Marta 3, 50139 Florence, Italy}
\author{Luigi Chisci}
\affiliation{Dipartimento di Ingegneria dell'Informazione, Universit\`{a} di Firenze,
Via S. Marta 3, 50139 Florence, Italy}


\begin{abstract}
Complex real-world phenomena are often modeled as  dynamical systems on networks. In many cases of interest, the spectrum of the underlying graph Laplacian sets the system stability and ultimately shapes the matter or information flow. This motivates devising suitable strategies, with rigorous mathematical foundation,  to generate Laplacian that possess prescribed spectra. In this paper, we show that a weighted Laplacians can be constructed so as to {\it exactly} realize a desired  {\it complex}  spectrum. The method configures as a non trivial generalization of existing recipes which assume the spectra to be real. Applications of the proposed technique to (i) a network of Stuart-Landau oscillators and (ii) to the Kuramoto model are discussed. Synchronization  can be enforced by assuming a properly engineered, signed and weighted, adjacency matrix to rule the pattern of pairing interactions.
\end{abstract}
\maketitle

\section{Introduction}

Complex networks play a role of paramount importance for a wide range of problems, of cross-disciplinary breadth. In several cases of interest, networks define the skeleton 
of pairwise interaction between coupled populations, families of homogeneous constituents anchored to a given node of the collections. The nature of  existing paired relationships 
between mutually entangled populations is encoded in the weight of the links,  that bridge adjacent nodes. Cooperative and competitive interactions can, in principle, be accommodated for by 
allowing for the weights to take positive or negative values. Local and remote non linear couplings shape the ensuing dynamics, and possibly steer the system towards a  stationary stable equilibrium  which is compatible with the assigned initial condition. The stability of the fixed points can be analyzed by studying the dynamical system in its linearized version.  
For reaction-diffusion systems defined on networks, the stability of the inspected equilibrium is ultimately dictated by the spectrum of the discrete Laplacian matrix \cite{newman,latora,barrat}. The eigenvalues of the Laplacian define, in fact, the support of the dispersion relation, the curve that sets the rate for the exponential growth of the imposed perturbation. More specifically,  external disturbances can be, in general, decomposed on the basis formed by the eigenvectors of the Laplacian operator. Each eigenvector defines an independent mode, which senses the web of intricate paths made accessible across the network: the perturbation can eventually develop, or, alternatively, fade away, along the selected direction, depending on the corresponding entry of the dispersion relation, as fixed by its associated eigenvalues. Stability is an attribute of paramount importance as it relates to resilience, the ability of a given system to oppose to external perturbations that would take it away from the existing equilibrium. Similarly, synchronization, a widespread phenomenon in distributed systems, can be enforced by properly adjusting the spectrum of the matrix which encodes for intertwined pairings. Based on the above,  it is therefore essential to devise suitably tailored recipes for generating networks, which display a prescribed Laplacian spectrum, compatible with the stability constraint \cite{cencetti1, cencetti2}. 

The problem of recovering a network from a set of assigned eigenvalues has been tackled in the literature both from an algorithmic \cite{ipsen,comellas,cvetkovic} and formal \cite{mckay,halbeisen} standpoints. In \cite{halbeisen}, a procedure is discussed to generate an undirected  and weighted graph from its spectrum.  The result extends beyond  
the well-known theorem of Botti and Merris \cite{botti} which states that the reconstruction of non-weighted graphs is, in general, impossible since almost all (non-weighted) trees share their spectrum with another non-isomorphic tree.  In \cite{motter}, a method is proposed to obtain a, directed or undirected, graph whose eigenvalues are constrained to match specific bounds, which ultimately reflect the nodes degrees, as well as the associated weights. In \cite{forrow}, a mathematically rigorous strategy is instead developed to yield weighted graphs which {\it exactly} realize any desired spectrum.  As discussed in \cite{forrow}, the method translates into an efficient approach to control the dynamics of various archetypal physical systems via suitably  designed Laplacian spectra. The results are however limited to undirected Laplacians, characterized by a real spectrum. The purpose of this paper is to expand beyond these lines, by proposing and testing a procedure which enables one to recover a signed Laplacian operator which displays a prescribed complex spectrum. Signed Laplacians are often used in the literature for applications which relate to social contagion, cluster synchronization or repulsive-attractive interactions \cite{bronski,altafini}. In engineering, they are often employed in modeling microgrids dynamics \cite{chen}.

 The paper is organized as follows. The first section is devoted to illustrating the devised method, focusing on the mathematical aspects. We then turn to discussing the implementation of the scheme and introducing the sparsification  algorithms that are run to cut unessential links.  In the subsequent section, we elaborate on the conditions that are to be met to generate a positively weigthed network. This discussion is carried out with reference to a specific setting. Then, we apply the newly introduced technique to the study of an ensemble made of coupled Stuart-Landau oscillators \cite{vanharten,aranson,garcamorales} and to (a simplified version of) the Kuramoto model \cite{kuramoto,strogatz}. Finally, in the last section, we sum up the contributions and provide concluding remarks.

%%%%%%%%%%%%%%%%

\section{A recipe to obtain a Laplacian with assigned complex eigenvalues.}

Consider a network made of $\Omega$ nodes and denote by $A$ the (weighted) adjacency matrix, where structural information is encoded. More precisely, the element $A_{ij}$ is different from zero when a directed link exists from $j$ to $i$. The entries of the matrix $A$ are real numbers and their signs reflect the specificity of the interaction at play: negative signs stand for inhibitory (or antagonistic) couplings, while positive entries point to excitatory (or cooperative) interaction. From the adjacency matrix, one can define its associated Laplacian operator. This is the matrix $L$, whose elements are $L_{ij}=A_{ij}-k_i \delta_{ij}$, where $k_i=\sum_j A_{ij}$ represents the natural extension of the concept of (incoming) connectivity to the case of a weighted network and $\delta_{ij}$ denotes the Kronecker delta.  

We shall here discuss a procedure to generate a Laplacian matrix, which displays a prescribed set of eigenvalues. As anticipated above,
we will focus in particular on directed Laplacians, which yields, in general, complex spectra. Concretely, we begin by introducing a collection of $\Omega=2N+1$ complex quantities defined as
\footnote{We shall assume all the eigenvalues but $0$ to be complex numbers. Let us remark that the method developed readily adapts to the case where also real eigenvalues are present. In this case, the eigenvectors associated to real eigenvalues are generated according to the prescriptions of \cite{forrow}. Eigenvectors linked to complex eigenvalues are instead assigned following the procedure outlined here.}

\begin{equation}
\label{eqn:spectrum}
\{\Lambda_i\}=\{\Lambda_1,\Lambda_2,\dots,\Lambda_{2N},\Lambda_{2N+1}=0\}
\end{equation}
The first $2N$ elements come in complex conjugate pairs and we set in particular $\Lambda_i=\Lambda_{i+N}^*$, $\forall$ $i=1,...,N$, where $(\cdot)^*$ stands for  complex conjugate. 
The aim of this section is to develop a rigorous procedure to  construct a directed (and weighted) graph $G$ with $2N+1$ vertices, whose associated Laplacian has $\{\Lambda_i\}$ 
for eigenvalues. Recall that $\{\Lambda_i\}$ contains the null element, since this latter is, by definition, a Laplacian's eigenvalue.  

The procedure that we are going to detail in what follows exploits the  eigenvalue decomposition of the Laplacian matrix. To this end we will seek to introduce a proper 
eigenvector basis such that 
\begin{equation}
\label{eqn:decomposition}
L=VDV^{-1}
\end{equation}
is a Laplacian. In Eq. \eqref{eqn:decomposition}, $D$ is a diagonal matrix where the sought Laplacian eigenvalues are stored. More specifically,  
$D_{ii}=\Lambda_{i-1}$ for $i=2,..,2N+1$ and $D_{11}=\Lambda_{2N+1}=0$. The problem is hence traced back to constructing $V$, whose columns are the right eigenvectors of $L$. We also recall that
rows of the inverse matrix $V^{-1}$ are the left eigenvectors of $L$. As outlined in \cite{cencetti3},  Laplacian (right and left) eigenvectors should satisfy a set of conditions:

\begin{enumerate} 
\item The columns of $V$, which refer to complex conjugate eigenvalues, must be complex conjugate too. 
\item The same condition holds for the rows of $V^{-1}$. 
\item Moreover, the columns of $V$ (resp. the rows of $V^{-1}$) corresponding to eigenvalues different from $0$ should sum up to zero. 
\item Finally, the right eigenvector relative to the null eigenvalue should be uniform (i.e. display identical components).
\end{enumerate}

In light of the above, we put forward for $V$ the following structure:  
\begin{equation}
\label{eqn:V}
V=
\begin{pmatrix}
c & v^T & {v^T}^* \\
cu & iU & -iU \\
cu & U & U
\end{pmatrix}
\end{equation}
where, $i$ stands for the imaginary unit, $U$ is an invertible $N\times N$ matrix having real entries, the vector $u=(1 \dots 1)^T$ has dimension $N\times 1$ and the vector $v$ is defined as
%\begin{equation}
%U=
%\begin{pmatrix}
%v_{11} & \dots & v_{1N} \\
%\vdots & \dots & \vdots  \\
%v_{N1} & \dots & v_{NN}
%\end{pmatrix}
%\end{equation}
%\begin{equation}
%v=
%\begin{pmatrix}
%-\sum_{j=1}^{N}v_{j1}(1+i) \\
%\vdots \\
%-\sum_{j=1}^{N}v_{jN}(1+i)
%\end{pmatrix}
%\end{equation}
\begin{equation}
\label{eqn:v}
v^T=-(1+i)u^TU
\end{equation}
The first column of $V$ is hence a uniform vector, corresponding to the eigenvector associated to the null eigenvalue. By construction, every other column sums up to zero, that is  
(\ref{eqn:v}) holds. In the following, we will write $D_{jj}=\alpha_j+i\beta_j$, which, in turn, implies $D_{j+N,j+N}=\alpha_j-i\beta_j$, for $j=2,\dots,N+1$. Here, $\alpha_j$ and $\beta_j$ are real quantities and respectively denote the real and imaginary parts of the $j$-th eigenvalue. To proceed further, one needs to determine the inverse of $V$.

To achieve this goal we begin by considering a generic matrix $W$, which satisfies the general constraints that are in place for $V^{-1}$. In formulae:
\begin{equation}
\label{eqn:W}
W=
\begin{pmatrix}
d & du^T & du^T \\
(1-i)w & S & -iS^* \\
(1+i)w^* & S^* & iS
\end{pmatrix}
\end{equation}
%where the vector
%\begin{equation}
%u=(1 \dots 1)^T
%\end{equation}
%satisfies
%\begin{equation}
%u^Tu=N
%\end{equation}
%and
%\begin{equation}
%uu^T=
%\begin{pmatrix}
%1 & \dots & 1 \\
%\vdots & \ddots & \vdots \\
%1 & \dots & 1
%\end{pmatrix}
%\end{equation}
where 
\begin{equation}
\label{eqn:w}
w=-\frac{1}{1-i}(Su-iS^*u)
\end{equation}

Note that the $j$th and $(N+j)$th rows of $W$, for $j=1,..,N+1$, are complex conjugated, as required. Moreover,  summing all the elements of each row (but the first) yields zero, a condition that the inverse of $V$ should meet, as anticipated above. Building on these premises, we shall here determine the unknown $S$, $w$ and $d$ so as to match the identity  $WV=I$, where $I$ stands for the $(2N+1) \times (2N+1)$ identity matrix. This implies, in turn, that $W \equiv V^{-1}$ due to the uniqueness of the inverse matrix.

A straightforward manipulation yields the following conditions for, respectively,  $d$ and $w$
%\begin{equation}
%\begin{cases}
%dc+dcu^Tu+dcu^Tu=1 \\
%v^T+iu^TU+u^TU=\mathbf{0} \\
%(1-i)w+Su-iS^*u=\mathbf{0} \\
%(1-i)wv^T+iSU-iS^*U=I \\
%(1-i)w{v^T}^*-iSU-iS^*U=\mathbf{0}
%\end{cases}
%\end{equation}
%that is
\begin{equation}
\begin{cases}
d=\frac{1}{c(2N+1)} \\
wv^T=i(S-iS^*)EU \\
w{v^T}^*=(S-iS^*)EU
\label{d_cond}
\end{cases}
\end{equation}
where use has been made of the identity $u^Tu=N$ and where 
\begin{equation}
\label{eqn:E}
E=uu^T=
\begin{pmatrix}
1 & \dots & 1 \\
\vdots & \ddots & \vdots \\
1 & \dots & 1
\end{pmatrix}
\end{equation}
The quantity $d$ is completely specified by the first of Eqs. (\ref{d_cond}) and solely depends on $c$ and  $N$, the size of the system. By making use of the identities \eqref{eqn:v} and \eqref{eqn:w}, one can progress in the analysis of the second and third conditions (\ref{d_cond}) to eventually get:
\begin{align}
SB+S^*A&=I \\
SA-S^*B&=0
\end{align}
where:
\begin{align}
A&=EU-iEU-iU \\
B&=iEU+EU+iU
\end{align}
It is, therefore, immediate to conclude that
\begin{align}
S&=(B+AB^{-1}A)^{-1}\\
S^*&=(B+AB^{-1}A)^{-1}AB^{-1}
\end{align}
The analysis can be pushed further to relate $S$ to matrices $U$ and $E$. The calculation, detailed in Appendix \ref{appS}, yields

\begin{equation}
\label{eqS}
S=-\frac{i}{2}U^{-1}\biggl[I-\frac{1+i}{1+2N}E\biggr]
\end{equation}

 The expression for $S^*$ can be immediately obtained by taking the complex conjugate of the above equation.
 In conclusion, the matrix $W$ defined in \eqref{eqn:W} is the inverse matrix of $V$, provided that $d$ and $S$ are respectively assigned as specified above.

Clearly,  matrix $L$ defined in \eqref{eqn:decomposition} has the desired spectrum \eqref{eqn:spectrum}. We should, however, prove that $L$ is a Laplacian. This amounts to showing that $L$ is a real  matrix, whose columns sum up to zero. The proof is given hereafter.

\vspace{0.5 truecm}

\textbf{Proposition.} Matrix $L$ is real. 
\\
From Eqs. \eqref{eqn:V} and \eqref{eqn:W}, one can readily compute the elements of $L$ via matrix products and, taking advantage of the block structures of $V$ and $W$,  prove that $L$ is real.  $\Re(\cdot)$ is introduced to represent the real part of $(\cdot)$.  The generic element $L_{st}$ can be written as:

\begin{eqnarray}
\label{eq:form1}
L_{st}=(VDW)_{st}&=&\sum_{k=1}^{2N+1}V_{sk}(DW)_{kt}=\sum_{k=1}^{2N+1}V_{sk}\sum_{k'=1}^{2N+1}D_{kk'}W_{k't}\\
&=& \sum_{k=2}^{2N+1}V_{sk}D_{kk}W_{kt}
\end{eqnarray}
due to the diagonal structure of the matrix $D$ and recalling that $D_{11}=0$. By making use of the specific 
form of $V$ and $W$, one gets:
\begin{eqnarray}
\label{eq:form2}
L_{st}&=&\sum_{k=2}^{N+1}V_{sk}D_{kk}W_{kt}+\sum_{k=N+2}^{2N+1}V_{sk}D_{kk}W_{kt}\\
&=&\sum_{k=2}^{N+1}V_{sk}D_{kk}W_{kt}+\sum_{k=2}^{N+1}(V_{sk}D_{kk}W_{kt})^*\\
&=&\sum_{k=2}^{N+1}2\Re(V_{sk}D_{kk}W_{kt})
\end{eqnarray} since $\alpha\beta+\alpha^*\beta^*=2\Re(\alpha\beta)$, for any complex numbers $\alpha$ and $\beta$. One can thus conclude that 
$L$, as generated by the above procedure, is real.\\

\vspace{0.5 truecm}

\textbf{Proposition.} Each column of $L$ sums up to zero.
\\
Because of the diagonal structure of $D$:
\begin{equation}
\sum_iL_{ij}=\sum_{il}V_{il}D_{ll}V^{-1}_{lj}
\end{equation}
Then:
\begin{equation}
\sum_iL_{ij}=\sum_lV^{-1}_{lj}D_{ll}\sum_iV_{il}=0
\end{equation}
since (i) $D_{11}=0$ and (ii) the components of all eigenvectors corresponding to non-null eigenvalues, sum up to zero. Notice that this result can also be proven by observing that the uniform vector $d\mathbf{1}$ is the left eigenvector corresponding to the null eigenvalue, that is
\begin{equation}
\label{eqn:sum_rows}
d\mathbf{1}L=0
\end{equation}
From \eqref{eqn:sum_rows}, it follows that $\sum_iL_{ij}=0$, for every $j$.\\

\vspace{0.5 truecm}

\textbf{Proposition.} $L$ is balanced. 
\\
We can also show that $\sum_jL_{ij}=0$ i.e. that the sum of all the elements of any given row $i$ returns zero.  According to 
\eqref{eqn:decomposition}, the first column of $V$ is the right eigenvector corresponding to eigenvalue $0$, namely
\begin{equation}
Lc\mathbf{1}=cL\mathbf{1}=0
\end{equation}
This implies, in turn, $\sum_jL_{ij}=0$ $\forall i$, which ends the proof. The Laplacian is hence balanced, as the sums on the rows and on the (corresponding) columns return the same result. 

From the generated Laplacian operator, one can readily calculate the adjacency matrix of the underlying network. In general, for any assigned spectrum, the recovered adjacency matrix is fully connected, meaning that there exists links connecting each pair of nodes. Notice that links are weighted and signed. The weights can be small or have a modest impact on the spectrum of the associated Laplacian. This motivates the implementation of a dedicated sparsification procedure, which seeks to remove unessential links, in terms of their reflection on the ensuing Laplacian spectrum. The next section is devoted to elaborating along these lines. 

\section{Examples and sparsification.}

\begin{figure}
\centering
\begin{tabular}{cc}
\includegraphics[scale=0.5]{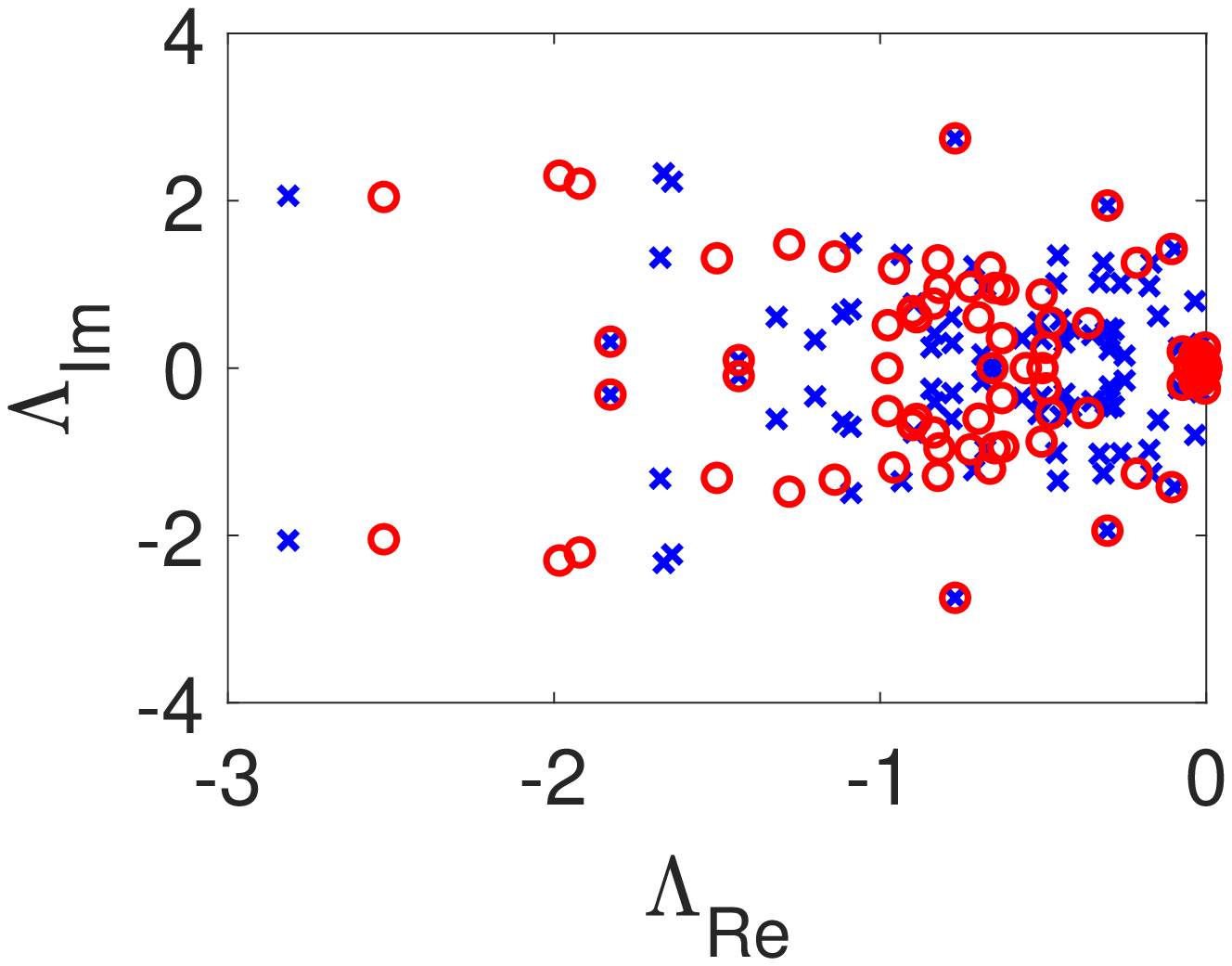} & 
\includegraphics[scale=0.5]{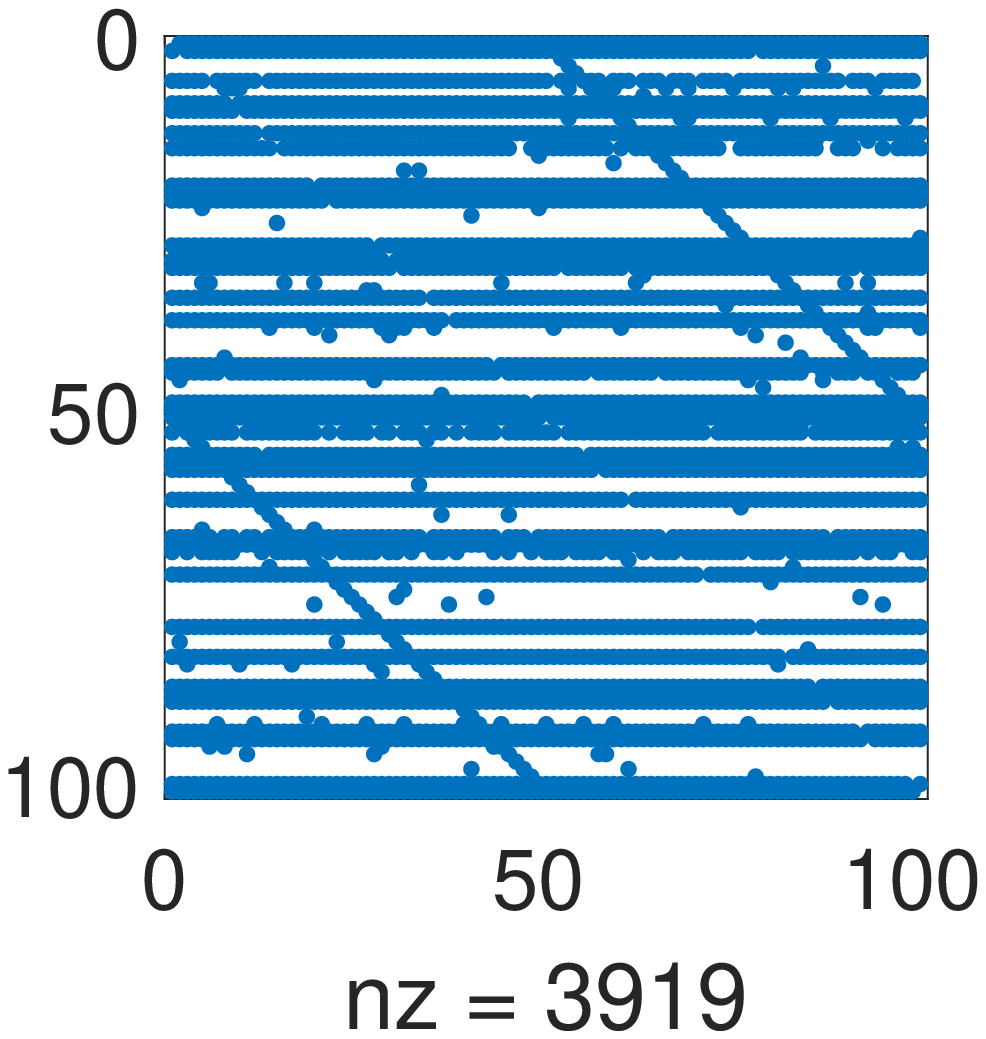} \\
(a) & ( b) \\
\includegraphics[scale=0.5]{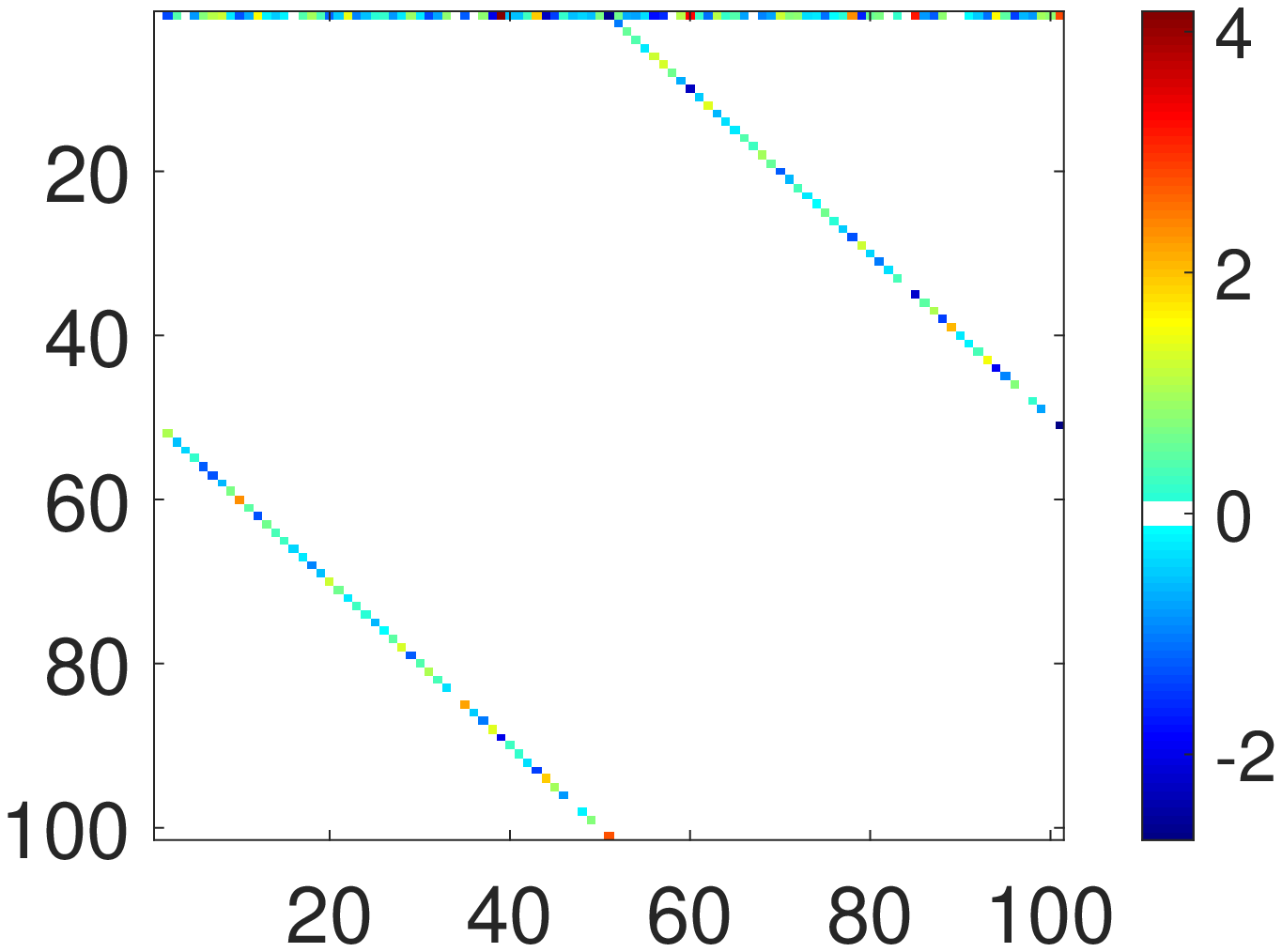} & 
\includegraphics[scale=0.5]{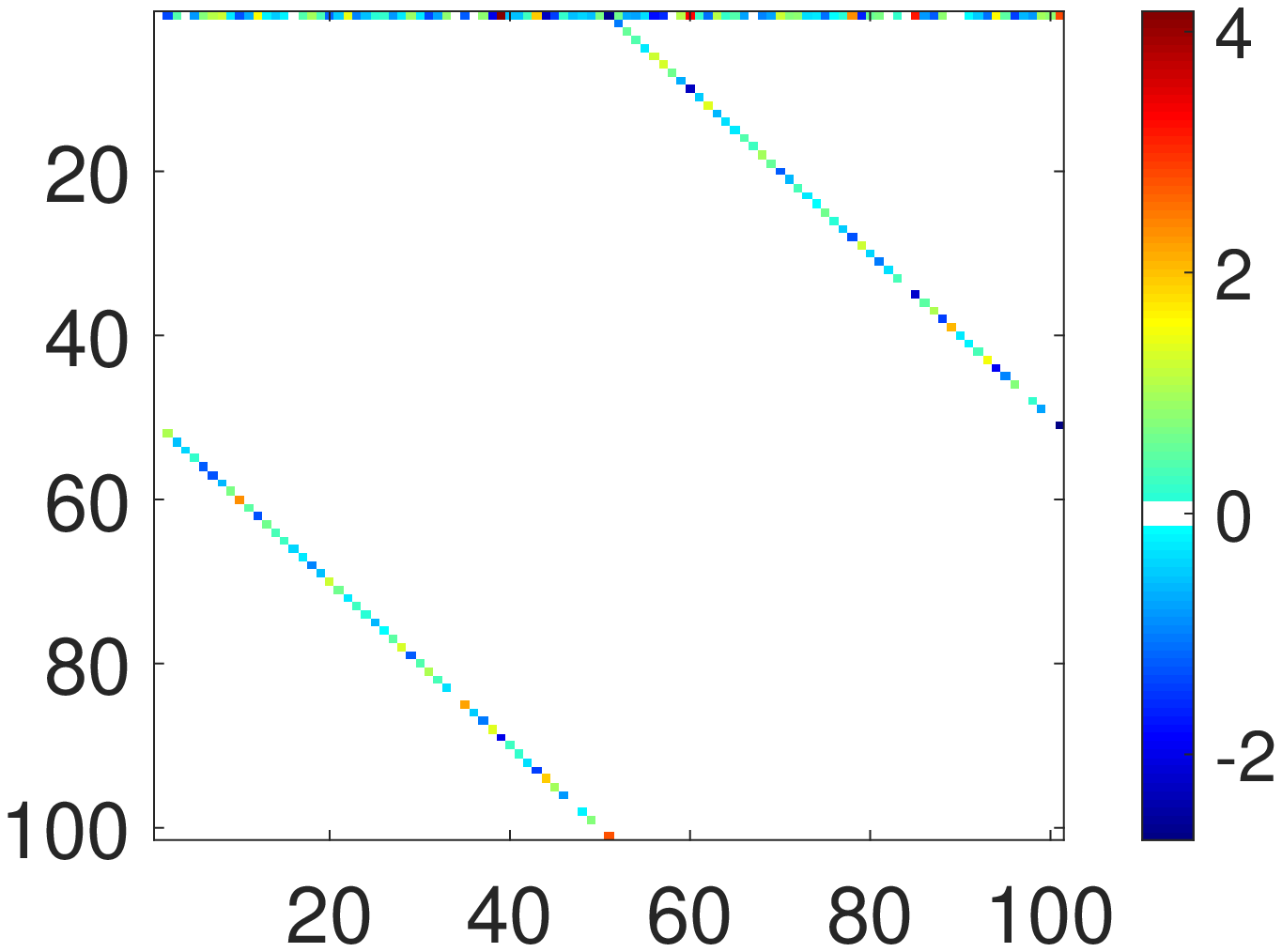} \\
(c) & ( d) \\
\end{tabular}
\caption{\it The recipe discussed in the main text is applied to generate a network of $\Omega=2N+1=101$ nodes, whose associated Laplacian displays the spectrum depicted (blue stars) in panel (a). The spectrum of the Laplacian obtained from the sparsified network is shown with red circles. We here follow the first sparsification recipe as illustrated in the main body of the paper. More specifically, we trim unessential links, chosen among those that bear very modest weights, while confining the Laplacian spectrum in a bounded domain, located in the negative portion of the complex plane. Here, $\sigma$ is $0.01$ and $\delta=0.5$. In panel (b), the sparsity pattern of the adjacency matrix obtained upon application of the sparsification algorithm is shown. The entries of the adjacency matrices, before and after the sparsification are respectively plotted, with an appropriate color-code, in panels (c) and (d). The main structure of the network is preserved upon application of the devised sparsification protocol.}
\label{f:fig1} 
\end{figure}

In this section we discuss a sparsification procedure, which aims at a posteriori simplifying the structure of  the recovered network. To this end, we begin by generating a network following the strategy outlined in the preceding section, and which yields an assigned spectrum for the associated Laplacian. The Laplacian spectrum that we seek to recover consists of $\Omega=2N+1$ complex entries, the eigenvalues, which are here confined in the left portion of the complex plane,  by setting $\Re(\Lambda_j)=\alpha_j<0$ for $j \ge 2$, see blue crosses in Fig. \ref{f:fig1} (a). This choice is somehow arbitrary, and ultimately amounts to enforce stability into a linear system of the form:

\begin{equation}
\label{lin_syst}
\frac{d x_i}{dt} = \sum_j L_{ij} x_j
\end{equation}
where $x_i$ is the $i$-th entry of the $\Omega$-dimensional state vector $x$. In the final part of the paper, we will turn to considering more complex scenarios where the stability of the examined dynamics is also influenced by local reaction terms.

The network that we obtain, following the scheme outlined in the preceding sections yielding a Laplacian with the prescribed spectrum, is in general fully connected. In other words, 
a weighted link exists between any pair of nodes. The weights of the link can be, in principle, very small and, as such, bear a modest imprint on the ensuing Laplacian spectrum. Motivated by this observation, we perform an a posteriori sparsification of the obtained network: this aims to identifying and then removing the finite subset of links that appear to have a modest impact on the eigenvalues of the associated Laplacian.

The first sparsification procedure that we have considered,  aims at removing unessential links while confining the spectrum of the Laplacian operator within a bounded region of the complex plane. More precisely, we focus on the links which display a weight in the range (-$\sigma$, $\sigma$), where $\sigma$ is a small, arbitrarily chosen, 
cut off. All links whose weight is smaller that $\sigma$ in absolute value are selected, in a random order. The selected link is removed and the modified Laplacian spectrum computed. 
Denote by $\tilde{\Lambda}_j$, for $j=2,...,2N+1$, the Laplacian eigenvalues obtained upon removal of the link. The change to the network arrangement becomes permanent, if 
$|min_j\left[\Re(\tilde{\Lambda}_j)\right]-min_j\left[\Re({\Lambda}_j)\right]| < \delta$ and $|max_j\left[\Im(\tilde{\Lambda}_j)\right]-max_j\left[\Im({\Lambda}_j)\right]|< \delta$, for $j=2,...,N$. Here $\Im(\cdot)$ stands for the imaginary part of $(\cdot)$ and $\delta$ is an arbitrary threshold which quantifies the amount of perturbation that is deemed acceptable for the problem at hand. As a further condition, we check that $\Re(\tilde{\Lambda}_j) <0$ for $j=2,...,2N+1$,  which, in turn, corresponds to preserving the stability of the linear system (\ref{lin_syst}). Clearly, the order of extraction of the links, which are candidate to be trimmed, matters. Different realizations of the procedure of progressive sparsification illustrated above might hence result in distinct final outcomes.
In Fig. \ref{f:fig1}(a), the eigenvalues obtained after the sparsification algorithm are  plotted (red circles) for one choice of the cutoff $\delta$.  The sparsity pattern of the adjacency
matrix obtained at the end of the above procedure is displayed in panel (b) of Fig. \ref{f:fig1}. In panels (c) and (d) of Fig. \ref{f:fig1} we plot, with an appropriate color code, the entries of the adjacency matrices, before and after the sparsification. Only weights which are significantly different from zero (see annexed colorbars) are displayed. As appreciated by  visual inspection, the  skeleton of the network is not altered by the applied sparsification.  To monitor how the eigenvalues get redistributed within the bounded domain to which they belong, we introduce the following indicators: 

\begin{equation}
I_x=\sum_{i=2}^{2N+1}(\beta_i)^2
\end{equation}
\begin{equation}
I_y=\sum_{i=2}^{2N+1}\biggl(\alpha_i -\frac{1}{2N+1}\sum_{j=2}^{2N+1}\alpha_j \biggr)^2 
\end{equation}

\begin{figure*}
\centering
\begin{tabular}{cc}
{\includegraphics[scale=0.4]{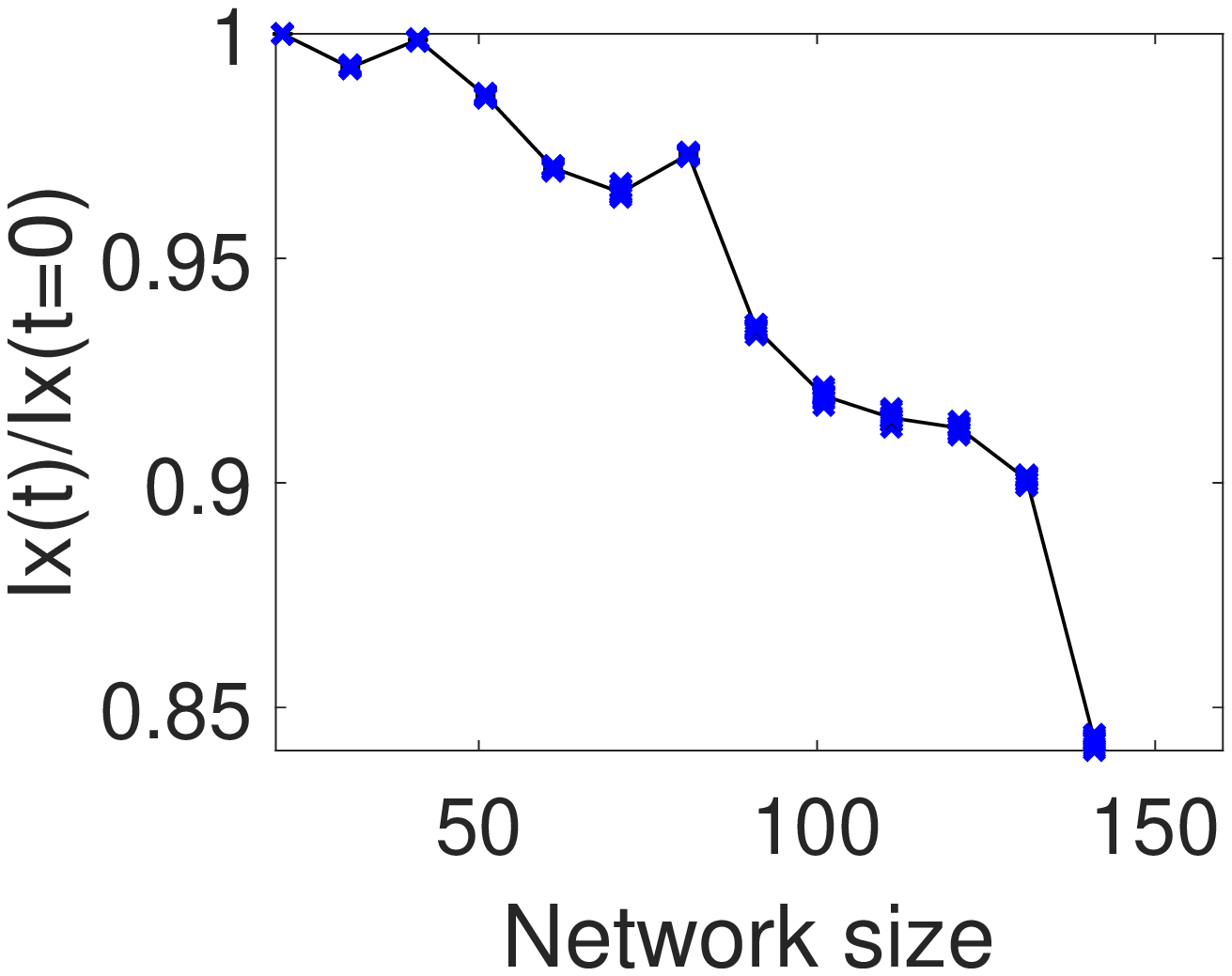}} &
{\includegraphics[scale=0.4]{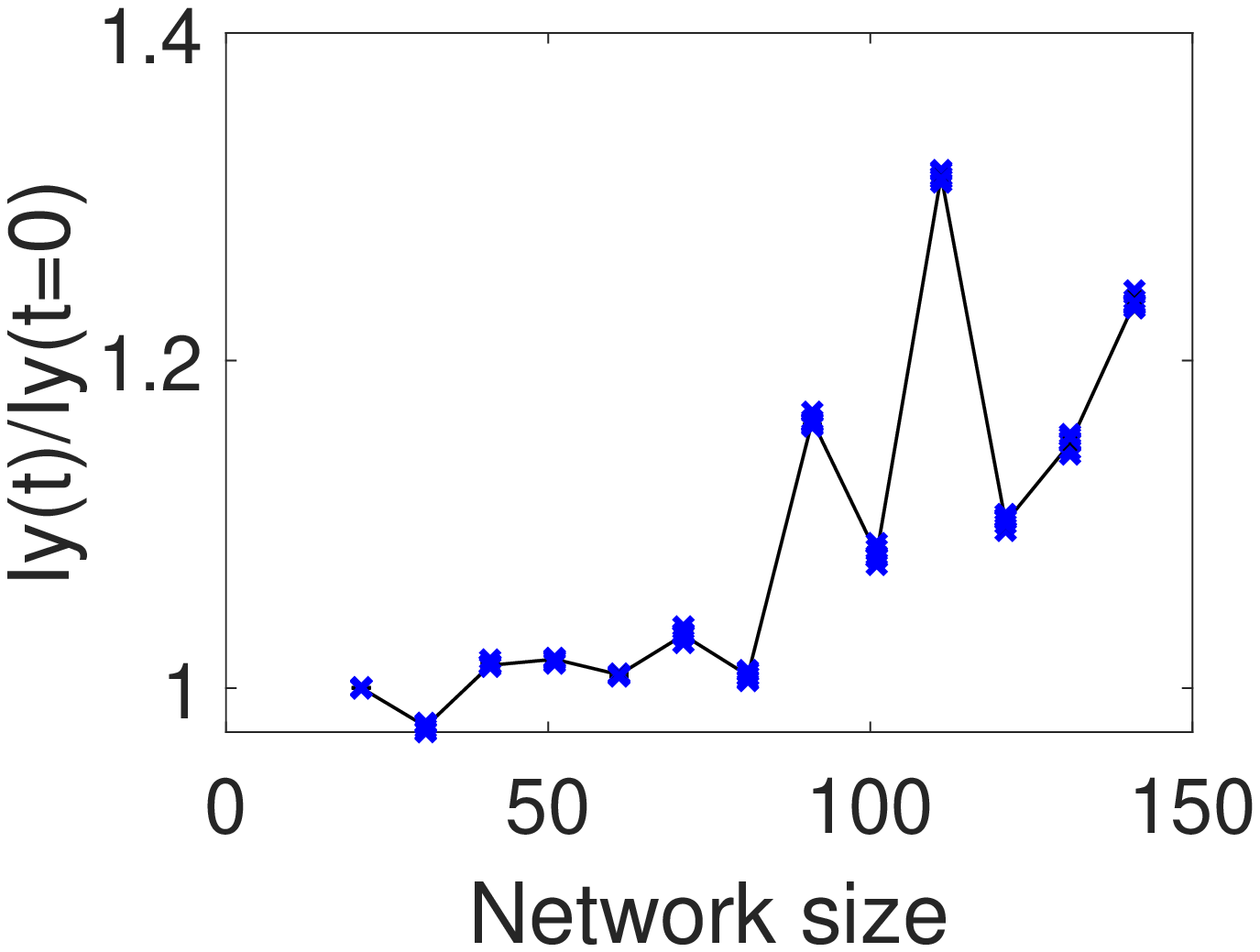}} \\
(a) & ( b) \\
\end{tabular}
\caption{\it Panel (a):  $I_x$ as measured at the end of the sparsification and normalized to the corresponding value, before the sparsification is plotted against $N$, the size of the explored network.  Panel (b): $I_y$, calculated after the  sparsification and normalized to the corresponding value, before the sparsification is depicted versus $N$. In both cases, blue symbols are computed, as the average over $15$ different realizations of the generated network (with the same given spectrum). For each value of $N$, the real and imaginary components of the eigenvalues are random number, drawn from a normal distribution. The solid line is a guide for the eye. Here, $\delta =0.5$ and 
$\sigma=0.01$.}
\label{f:fig2}
\end{figure*}

The quantity $I_x$ measures the dispersion along the imaginary axis, by weighting the squared distance of each eigenvalue from the horizontal axis. Conversely,   $I_y$ reflects the scattering 
of the eigenvalues about their mean, in the direction of the real axis.  In Fig. \ref{f:fig2}, $I_x$ and $I_y$, normalized to their respective values obtained before application of the sparsification algorithm, are shown against $N$, an indicator of the size of the generated networks. The sparsification procedure shrinks the eigenvalues in the $x$-direction, while the opposite tendency is observed for the distribution along the $y$-direction.
 
\begin{figure*}
\centering
\begin{tabular}{cc}
{\includegraphics[scale=0.4]{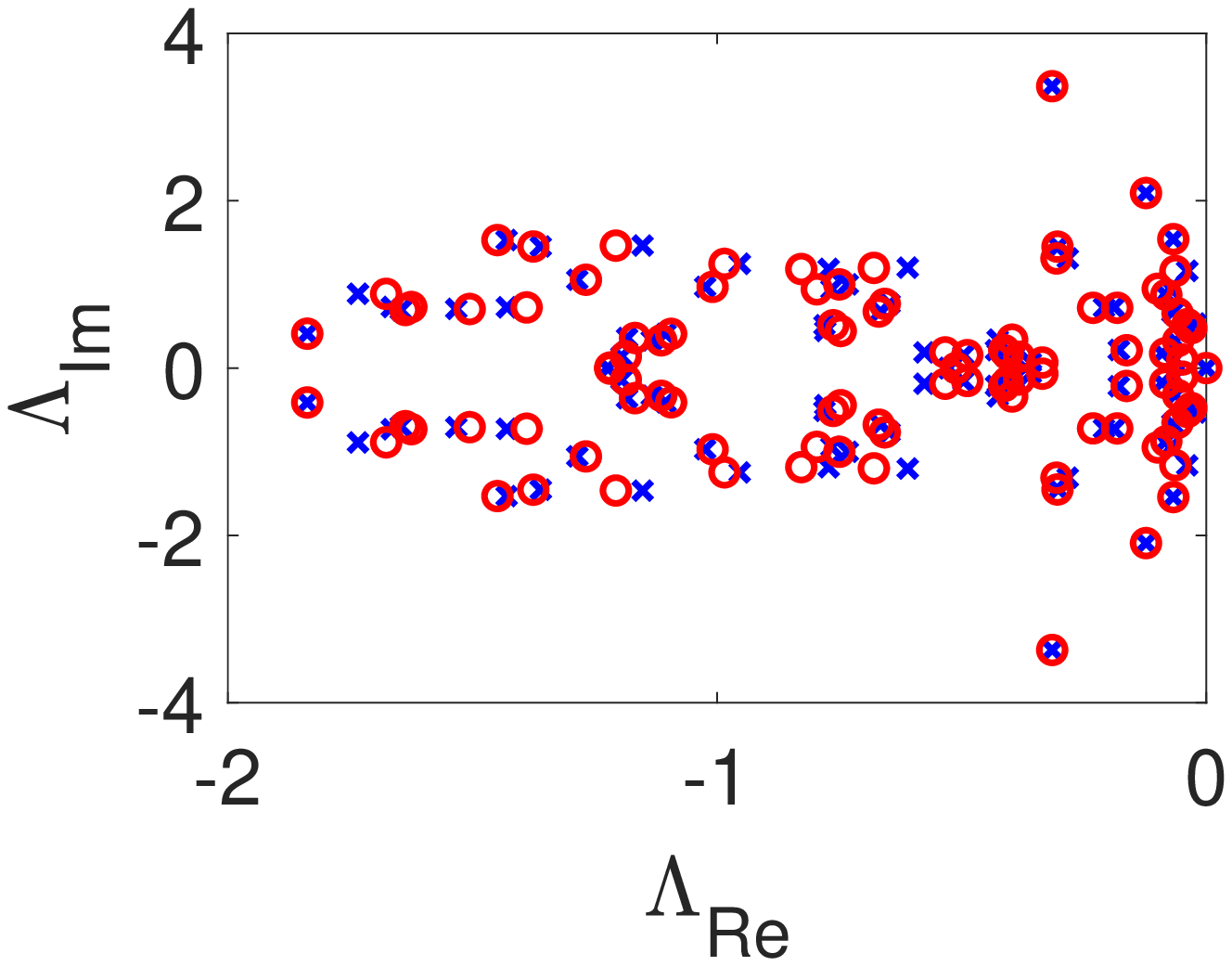}} &
{\includegraphics[scale=0.4]{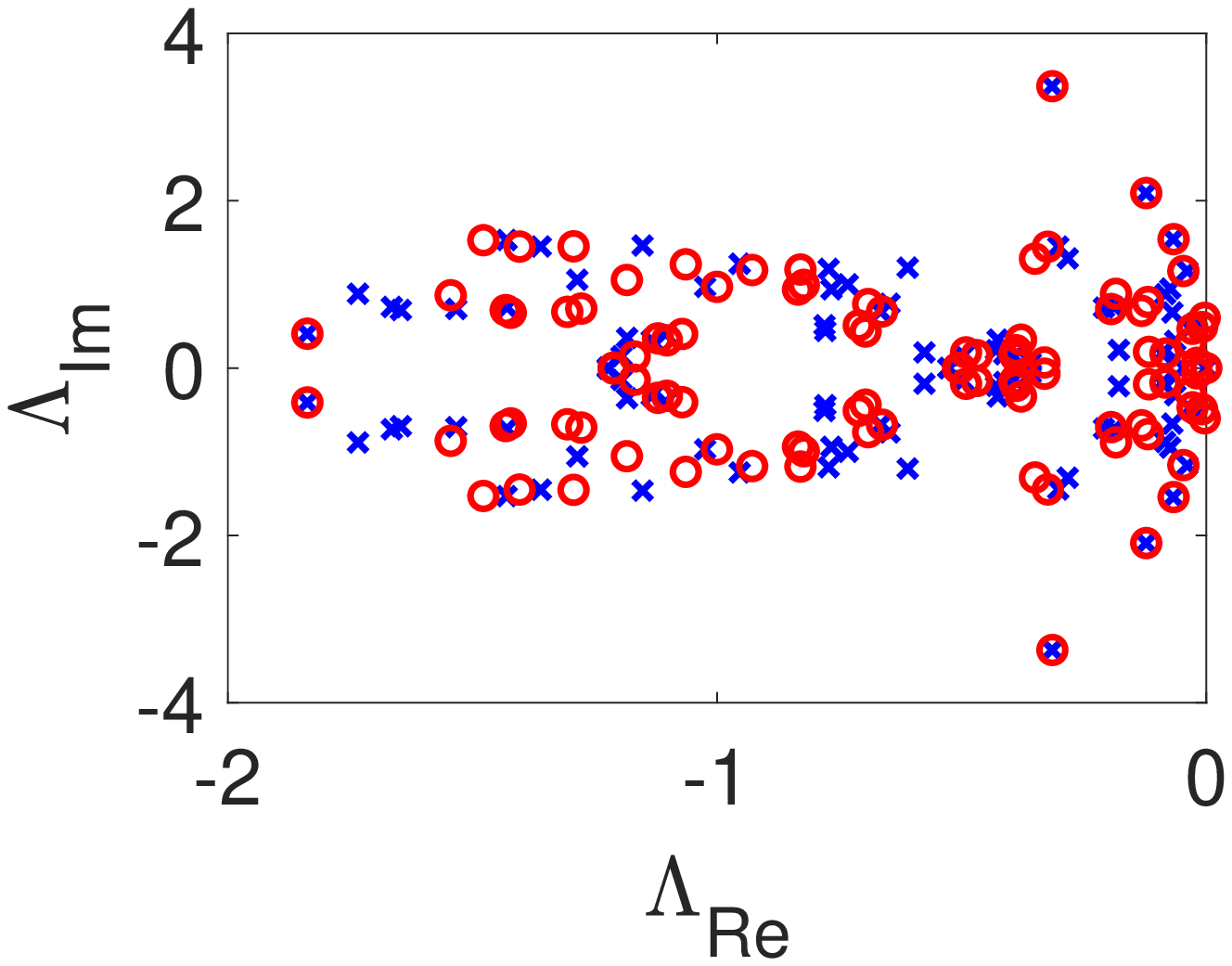}} \\
\\
(a) & ( b) \\
\end{tabular}
\caption{\it Effect of the second method of sparsification on a network made of $\Omega=2N+1=101$ nodes.  The original spectrum is plotted with (blue) stars. The modified one with (red) circles. Panel (a) refers to $\delta=0.07$  while panel (b) to $\delta=0.2$. Here, $\sigma=0.01$.}
\label{f:fig3}
\end{figure*}

\begin{figure*}
\centering
\includegraphics[scale=0.5]{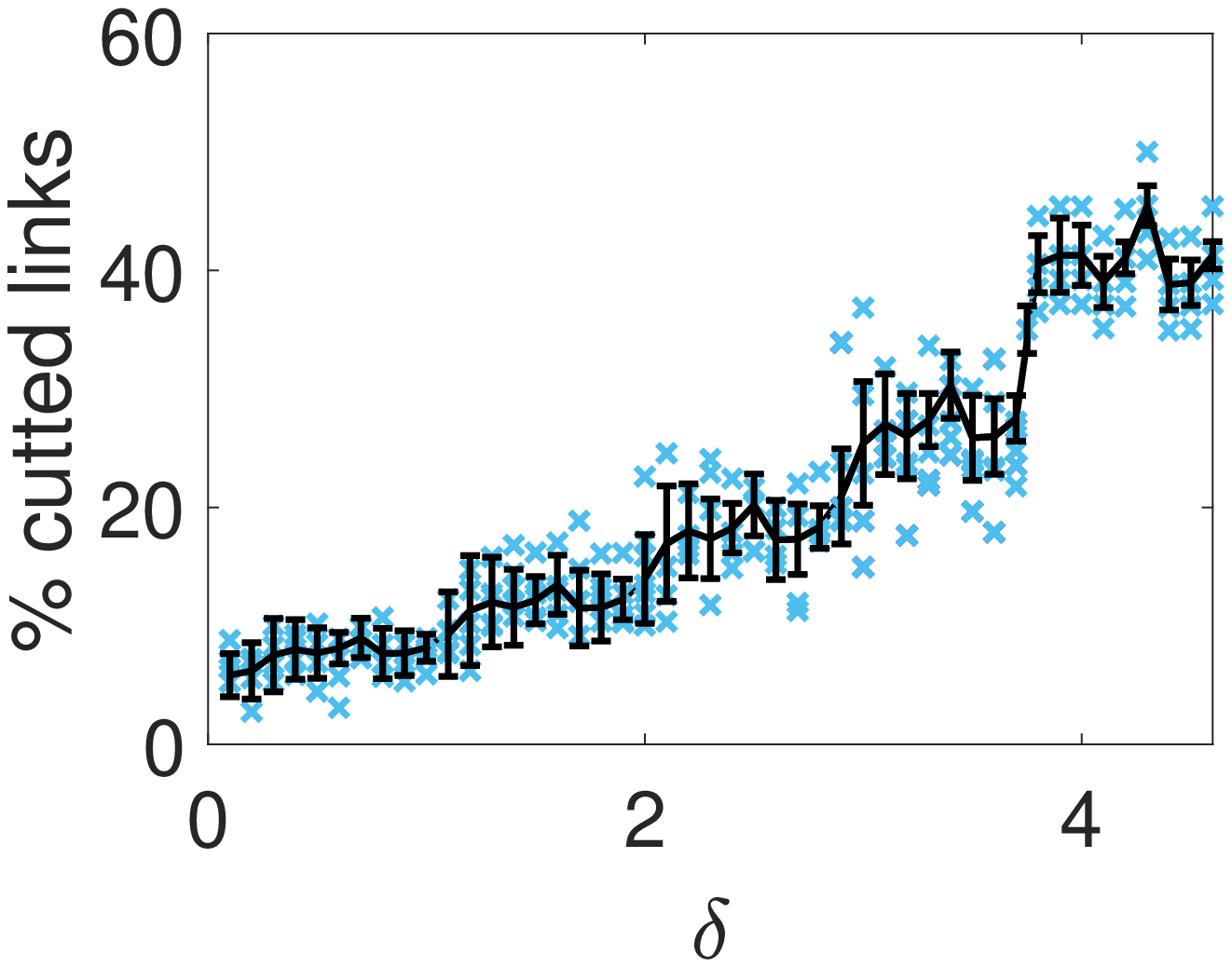}
\caption{\it The percentage of links that are cut against the allowed perturbation $\delta$.  Blue symbols refer to $5$ independent 
realizations, for each value of $\delta$. The black line goes through the average values, computed from the collection of independent runs, at fixed $\delta$. Here, $N=50$ ($\Omega=2N+1=101$) and $\sigma=0.01$.}
\label{f:fig4}
\end{figure*}

The second sparsification method implements a more stringent constraint. Just like before, we select the links with weights in the range (-$\sigma$, $\sigma$), where $\sigma$ acts as a small threshold amount. Unlike with the former case, we now eliminate the selected link only if the change produced on the modulus of {\it each} of the $N$ eigenvalues is smaller than $\delta$, namely if  
$|{\Lambda}_j- \tilde{\Lambda}_j| < \delta$, for $j=2,...,2 N+1$. In Fig. \ref{f:fig3}, the eigenvalues obtained after the sparsification algorithm are  plotted (red circles) for two choices of the cutoff $\delta$.  The number of links that can be effectively removed grows with $\delta$, the size of the allowed perturbation, as clearly demonstrated in Fig. \ref{f:fig4}.

Summing up, we have developed and tested a procedure to generate a network which returns an associated Laplacian matrix with a prescribed complex spectrum. The weighted network obtained following the above procedure is, in general, fully connected. Dedicated sparsification strategies can be applied to remove the links which carry a small weight, and bear a modest 
imprint on the ensuing Laplacian spectrum. In the following, we will consider a specific setting of the aforementioned generation scheme, which makes it possible for the Laplacian elements to be computed analytically.

%%%%%%%%%%%%%%%%%%%%%%%%%

\section{Focusing on the special case $U=qI$}

In the previous sections we described a general method to generate a  Laplacian matrix which displays a designated spectrum. The method assumes a generic matrix $U$, which can be 
randomly assigned. In the following, we will focus on the specific case where $U$ is proportional to the identity matrix and progress with the analytic characterization of the obtained Laplacian.
As we shall argue in the following, working in this framework allows us to derive a set of closed conditions for constraining the weights of the underlying network to strictly positive values.  
To proceed in this direction we set:

\begin{equation}
U=qI
\end{equation}
where $I$ stands for the identity matrix and $q$ is  scalar.

%In general, whatever is the form of $U$, due to the structure of the matrix $V$, the node enumerate as the first one is higly connected to the other ones (see Fig \ref{fig:sparsification1}). To %circumvent this fact one can think to permute in an appropriate way the elements of the columns of $V$, starting for example from column two (or three) and every second column with the %same permutation rule.
%For the specific case $U=I$, the permutation of the elements in $V$ could involve a problem for the inversion of the matrix $V$. To circumvent this impediment we can think to extract $U$ from %a normal distribution centered in $I$ in the sense that the diagonal elements are extracted from a normal distribution centered in $1$ and the off-diagonal elements from a normal distribution %centered in $0$, both with the same variance $\sigma_U^2$. In this way we avoid potential numerical issues. At the same time the Laplacian that results from this choice is not so different from %the one derived using $U=I$. We see that smaller is $\sigma_U^2$ higher is the peak in the distribution of the entries of $L$. We observe also that the elements of the adjacency matrix do not %follow a normal distribution with variance $\sigma_U^2$, as well.

A straightforward calculation returns the following expression for matrix $S$:
\begin{equation}
\label{eq:structure_W}
S=
\begin{pmatrix}
a+ib & a-ia & \dots & a-ia \\
a-ia & a+ib & \ddots & a-ia \\
\dots & \ddots & \ddots & \vdots \\
a-ia & \ddots & \ddots & a+ib
\end{pmatrix}
\end{equation}
while $w$ is a uniform vector with identical entries equal to $Na+((N-1)a-b)i$
and the quantities $d,a,b$ are specified by
\begin{align}
d&=\frac{1}{(2N+1)c} \\
a&=\frac{-1}{2(2N+1)q} \\
b&=2Na
\end{align}

From equations \eqref{eq:form1}, \eqref{eq:form2} one can obtain a closed expression for each element of the Laplacian, as function of the eigenvalues. The interested reader can find the detailed computations in Appendix~\ref{app:contoL}. In the following the final  formulae are reported. 

The diagonal elements satisfy:
\begin{eqnarray}
\label{L11}
L_{11}&=&\frac{2}{2N+1}\sum_{k=2}^{N+1}\alpha_k\\
L_{ss}&=&\frac{2N\alpha_s+\beta_s}{2N+1} \quad s=2,\dots ,N+1\\
L_{ss}&=&\frac{2N\alpha_s-\beta_s}{2N+1}\quad s=N+2,\dots ,2N+1\, ,
\end{eqnarray}
where use has been made of the identity $\Re(D_{kk})= \alpha_k$. The first row and column are given by:
\begin{eqnarray}
L_{1t}&=&\frac{1-2N}{1+2N}\alpha_t-\beta_t+\frac{2}{2N+1}\sum_{k=2,k\neq t}^{N+1}\alpha_k\quad t=2,\dots,N+1\\
L_{1t}&=&\frac{1-2N}{1+2N}\alpha_t+\beta_t+\frac{2}{2N+1}\sum_{k=2,k\neq t-N}^{N+1}\alpha_k \quad t=N+2,\dots,2N+1\\
L_{t1}&=&\frac{-\alpha_t+\beta_t}{2N+1}\quad t=2,\dots,N+1\\
L_{t1}&=&\frac{-\alpha_t-\beta_t}{2N+1}\quad t=N+2,\dots,2N+1\, .
\end{eqnarray}
while the remaining elements read:
\begin{eqnarray}
L_{st}&=&\frac{-\alpha_s+\beta_s}{2N+1}\quad s=2,\dots,N+1 \text{ and $t=N+2,\dots,2N+1$ with $t\neq s-N$}\\
L_{st}&=&\frac{-\alpha_s-\beta_s}{2N+1}\quad s=N+2,\dots ,2N+1 \text{ and $t=2,\dots,N+2$ with $t\neq s-N$}\\
L_{s,s+N}&=&\frac{-\alpha_s-2N\beta_s}{2N+1}\quad s=2,\dots,N+1\\
L_{s,s-N}&=&\frac{-\alpha_s+2N\beta_s}{2N+1}\quad s=N+2,\dots,2N+1\\
L_{st}&=&\frac{-\alpha_s+\beta_s}{2N+1} t,s=2,\dots,N+1 \text{ and $s\neq t$}\\
L_{st}&=&\frac{-\alpha_s-\beta_s}{2N+1}s,t=N+2,\dots,2N+1 \text{ and $s\neq t$}
\end{eqnarray}

\subsection{Controlling the sign of non-diagonal Laplacian entries.}

The Laplacian matrix obtained with the procedure illustrated above has both positive and negative entries. Signed Laplacians are often used in consensus problems, where negative weights model antagonistic interactions. In other contexts, when e.g. the Laplacian is stemming from diffusive interactions, non diagonal entries are constrained to positive values. In the following, we will provide a set of necessary conditions for the assigned spectrum to eventually yield a Laplacian with positive extra diagonal elements, $L_{ij}>0$ for $i \ne j$. Clearly, $L_{ii}<0$, as summing on the rows should return zero. The underlying network, therefore, displays positive weights, as its adjacency matrix is basically obtained from the Laplacian matrix by replacing the diagonal elements with zeros.

Further, we will set  $\Re(\Lambda_j)=\alpha_j<0$ for $j \ge 2$, an assumption which corresponds to dealing with a stable linear system of the type given in Eq. (\ref{lin_syst}). This requirement immediately yields $L_{11}<0$, as it follows from relation (\ref{L11}). We will also operate in the setting analyzed above, i.e. assuming  $U=qI$.  The obtained expressions for the Laplacian elements allow to recast the sought conditions on their signs as: 
%\begin{equation}
%\begin{cases}
%\sum_{k=2}^{N+1}\alpha_k<0 \\
%2N\alpha_s+\beta_s<0  \\
%2N\alpha_s-\beta_s<0  \\
%\frac{1-2N}{2N+1}\alpha_t-\beta_t+\frac{2}{2N+1}\sum_{k=2,k\neq t}^{N+1}\alpha_k>0 \\
%\frac{1-2N}{2N+1}\alpha_t+\beta_t+\frac{2}{2N+1}\sum_{k=2,k \neq t}^{N+1}\alpha_k>0  \\
%-\alpha_s+\beta_s>0  \\
%-\alpha_s-\beta_s>0  \\
%-\alpha_s-2N\beta_s>0  \\
%-\alpha_s+2N\beta_s>0 
%\end{cases}
%\end{equation}
%The system can be rewritten as

\begin{equation}
\begin{cases}
\label{eqn:sistema}
2N\alpha_t<\beta_t<-2N\alpha_t  \\
\beta_t>-\frac{1-2N}{2N+1}\alpha_t-\frac{2}{2N+1}\sum_{k=2,k\neq t}^{N+1}\alpha_k \\
\beta_t<\frac{1-2N}{2N+1}\alpha_t+\frac{2}{2N+1}\sum_{k=2,k\neq t}^{N+1}\alpha_k  \\
\alpha_t<\beta_t<-\alpha_t  \\
\frac{\alpha_t}{2N}<\beta_t<-\frac{\alpha_t}{2N}  \\
\end{cases}
\end{equation}
where the inequalities hold for $t=2,\dots,N+1$.

The above conditions can be simplified, after some algebraic manipulations, to return
\begin{equation}
\label{eqn:condition_alpha1}
\frac{4N}{4N^2-1}\sum_k\alpha_k<\alpha_t<\frac{2}{2N+1}\sum_k\alpha_k
\end{equation}
and
\begin{equation}
\label{eqn:condition_beta1}
\alpha_t-\frac{2}{2N+1}\sum_k\alpha_k<\beta_t<\\
-\alpha_t+\frac{2}{2N+1}\sum_k\alpha_k\, .
\end{equation}
The interested reader can access the detailed steps in Appendix~\ref{app:positL}.

Assume that the assigned Laplacian spectrum matches the above conditions, while having $\alpha_k<0$ for $k>2$. Then the Laplacian matrix obtained with the above procedure 
, with $U=qI$, displays positive non diagonal entries.

\begin{figure*}
\centering
\begin{tabular}{ccc}
{\includegraphics[scale=0.35]{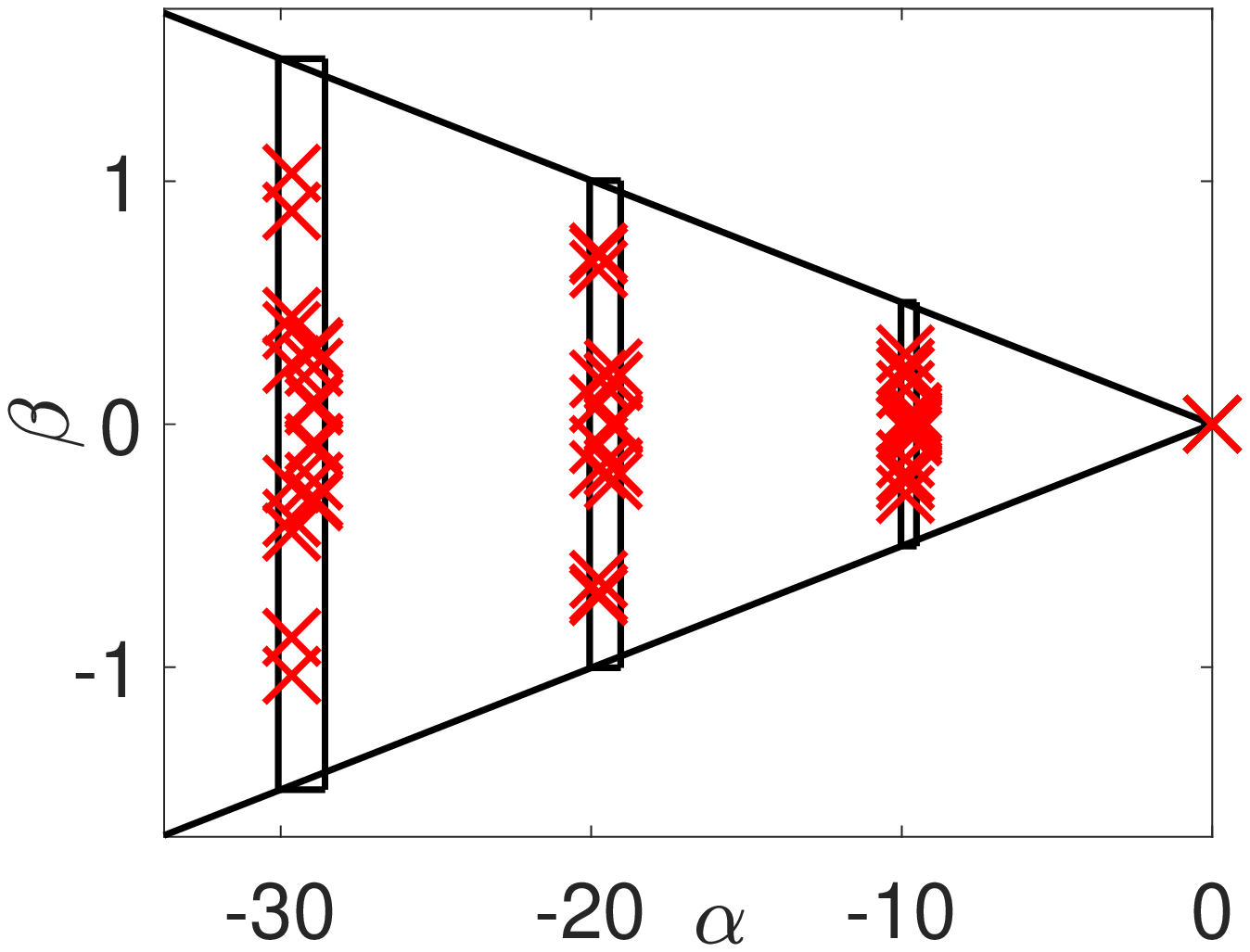}} &
{\includegraphics[scale=0.35]{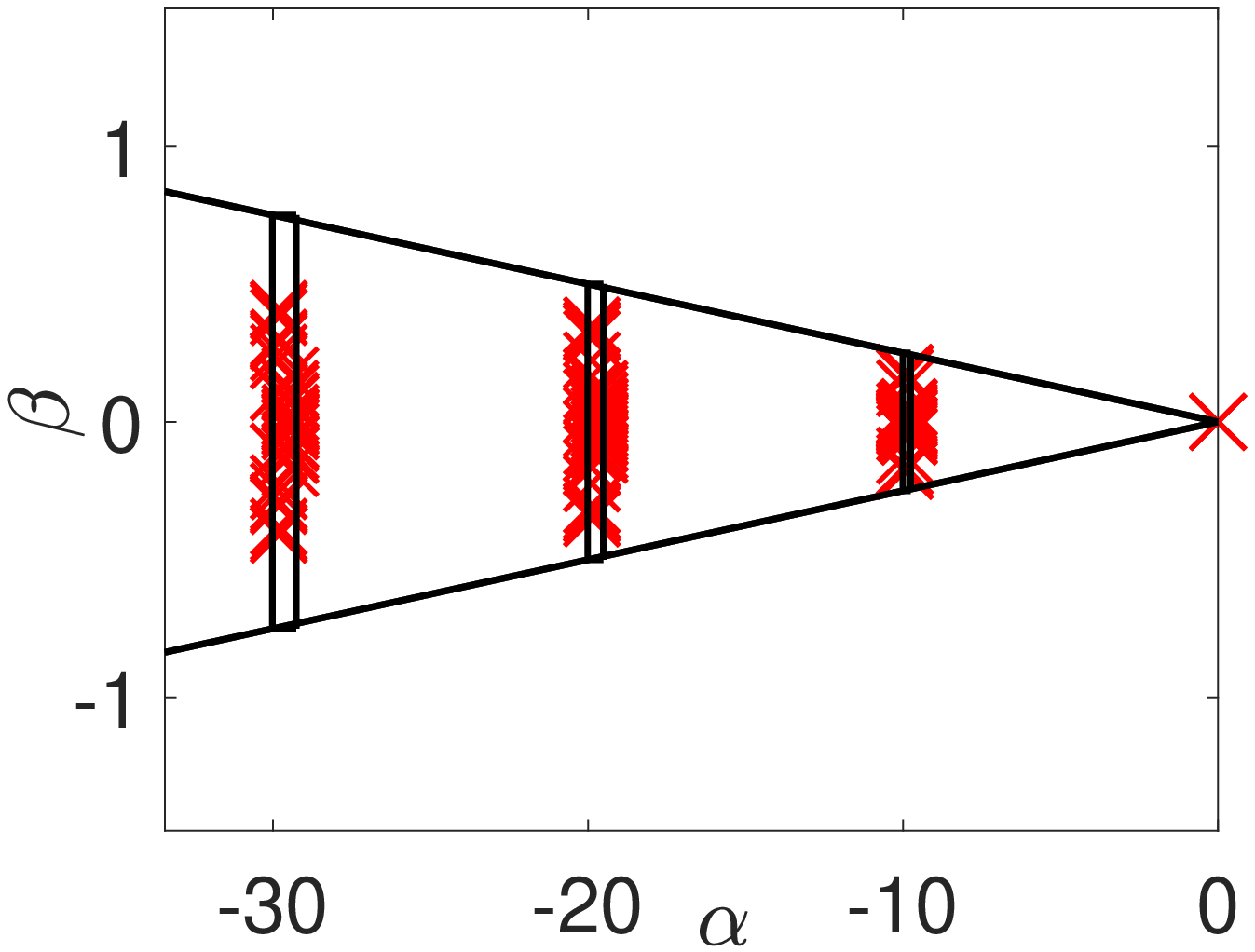}} &
{\includegraphics[scale=0.35]{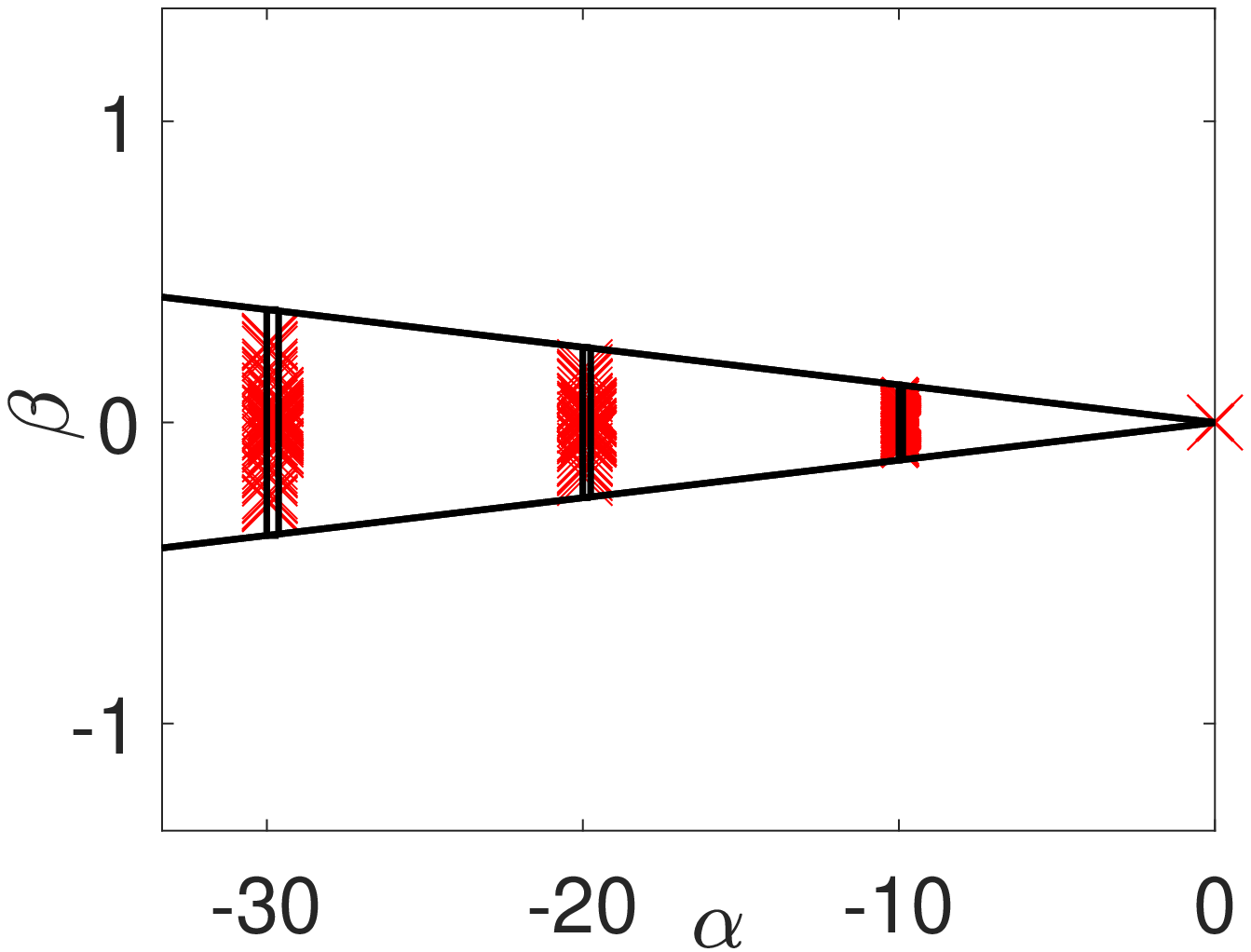}} \\
(a) & ( b) & (c) \\
\end{tabular}
\caption{\it Examples of discrete spectra (red symbols) which yield a Laplacian with positive non diagonal entries. The rectangular boxes identify the region where eigenvalue should fall for the ensuing network to display positive weights and are traced according to the conditions derived in the main body of the paper.  Here, $\Omega=2N+1=21$ (a), $\Omega=2N+1=41$ (b) and $\Omega=2N+1=81$ (c).}
\label{fig:degeneracy}
\end{figure*}

Let us explore the consequences of conditions (\ref{eqn:condition_alpha1}) and (\ref{eqn:condition_beta1}). To this end, introduce  $\bar{\alpha}$, the average of the non-negative real parts of the Laplacian eigenvalues, i.e. $\bar{\alpha}=(\sum_k \alpha_k)/(2N)$. A straightforward analysis allows us to conclude that
the generated Laplacian returns positive non-diagonal elements, if the assigned non trivial eigenvalues (i.e. $\Lambda_k$, with $k>1$) fall in a bounded rectangular domain of the complex plane. More specifically, the rectangular region is symmetric, with respect to the horizontal (real) axis, and extends along the vertical direction (imaginary axis) of $\pm 2N\bar{\alpha}/(4N^2-1)$. The rectangle is completed by two vertical sides, positioned at $ \frac{4N^2}{4N^2-1}\bar{\alpha}$ and $\quad \frac{2N}{2N+1}\bar{\alpha}$.  Working at fixed $N$, the larger is $|\bar{\alpha}|$ the more extended is the rectangle along the vertical direction. Conversely, when making $N$ larger, the rectangle shrinks in the horizontal direction and becomes eventually degenerate for $N \rightarrow \infty$. In other words, for large values of $N$, eigenvalues should align on a vertical segment positioned at  $\bar{\alpha}$, and whose extension increases linearly with $|\bar{\alpha}|$ (while decreasing with $N$). This is shown in Fig. \ref{fig:degeneracy}, where three different spectra are depicted (for different choices of $\bar{\alpha}$) which yield a Laplacian with positive off-diagonal elements. 

In the following section, we will apply the method of Laplacian generation here discussed to the study of two prototypical examples of dynamical systems on networks.

\section{Selected applications}

In this Section we consider two different models of interacting oscillators, defined on a network. In both cases, the coupling between individual oscillators is implemented via a 
discrete Laplacian operator, which reflects the specific network arrangement. It will be shown that a suitable network arrangement can be a priori established, building on the procedure illustrated above, so as to make the inspected systems stable against external perturbations.  

\subsection{Coupled Stuart-Landau oscillators}

Consider an ensemble made of $2N+1$ nonlinear oscillators and denote by $W_i$ their associated complex amplitude. We assume the oscillators to be mutually coupled via a diffusive-like interaction which is mathematically modeled by a discrete Laplacian operator. Each oscillator obeys a complex Stuart-Landau equation. The dynamics of the system can be cast in the form:
\begin{equation}
\label{GL}
\frac{d}{dt}W_j=W_j-(1+ic_2)|W_j|^2W_j+(1+ic_1)K\sum_kL_{jk}W_k
\end{equation}
where $c_1$ and $c_2$ are real parameters. The index $j$ runs from $1$ to $2N+1$, the total number of oscillators. Here, $K$ is a suitable parameter setting the coupling strength. Without loss of generality, in what follows it is assumed that $K=1$. $L_{ij}=A_{ij}-k_i \delta_{ij}$ is the Laplacian, $A_{ij}$ is the generic entry of the directed and weighted adjacency matrix $A$ and $k_i=\sum_j A_{ij}$.

\begin{figure*}
\centering
\begin{tabular}{ccc}
{\includegraphics[scale=0.35]{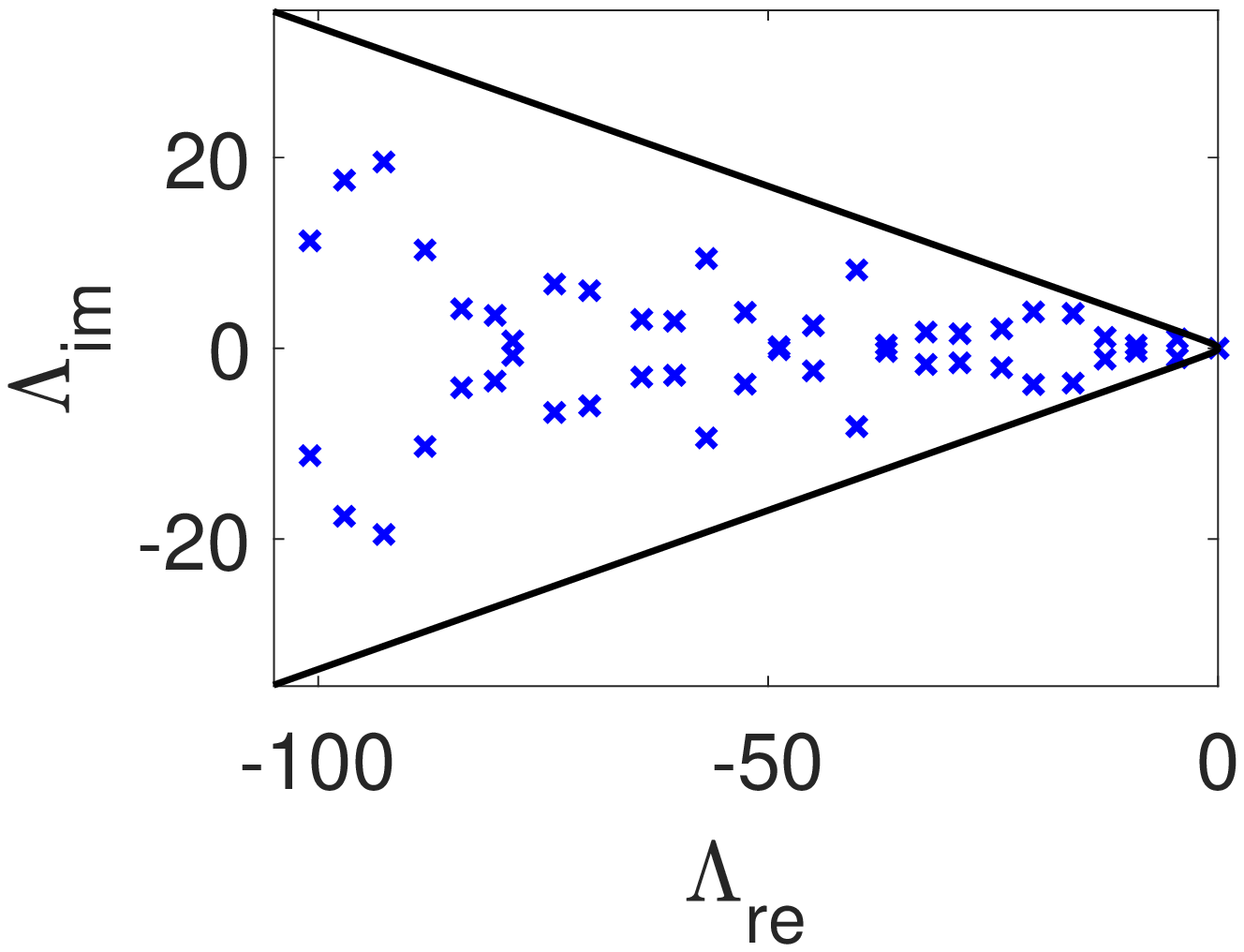}} &
{\includegraphics[scale=0.35]{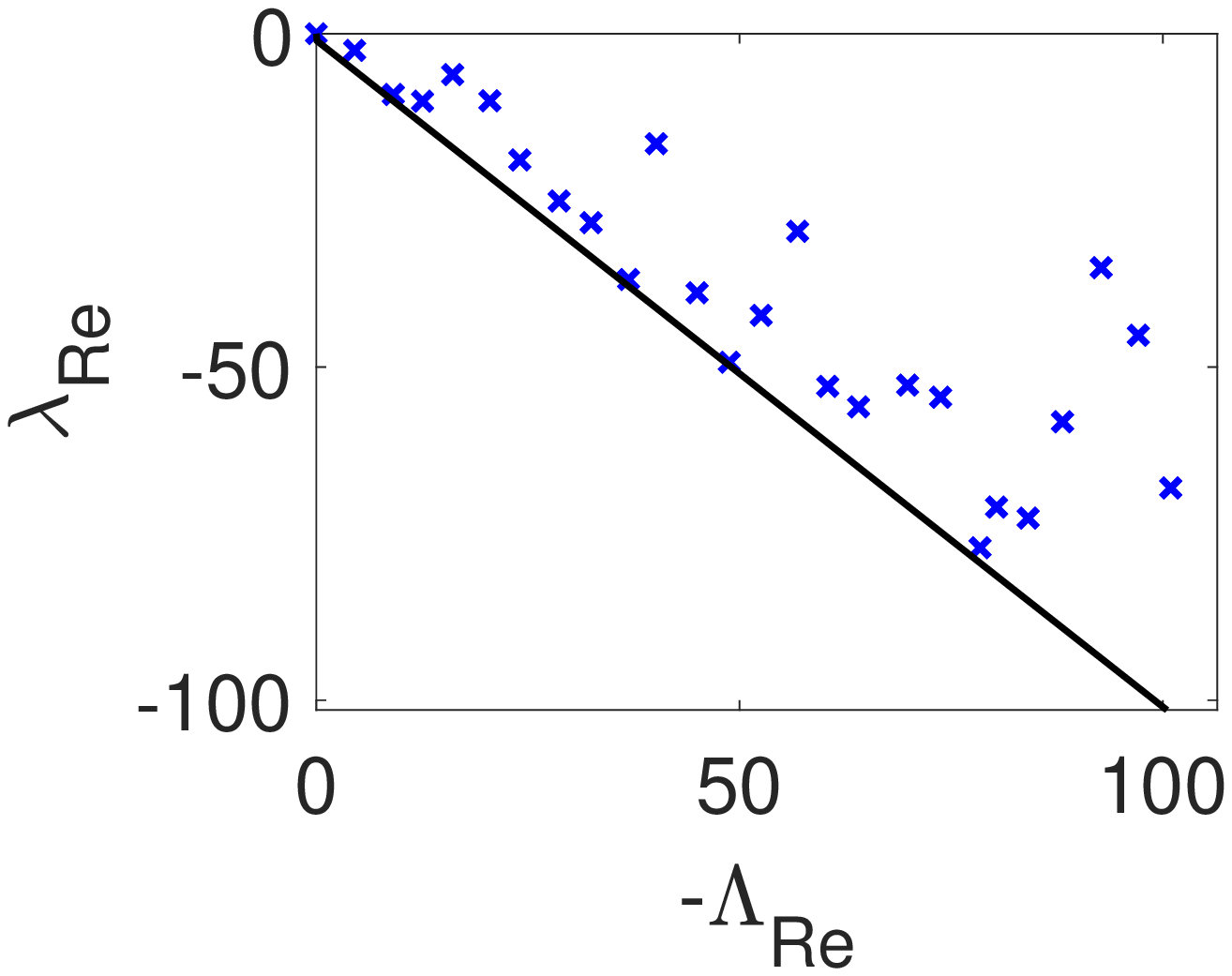}} &
{\includegraphics[scale=0.35]{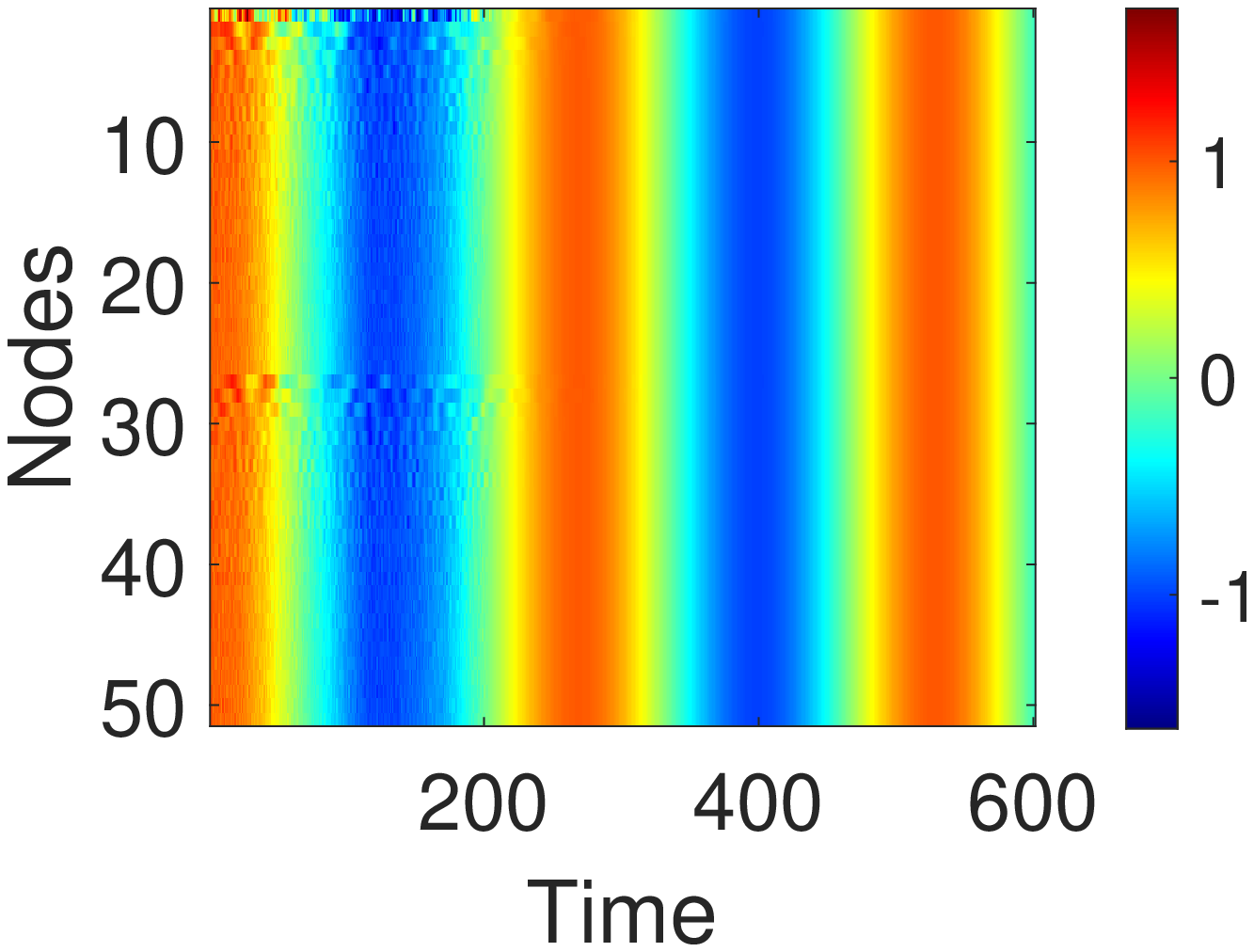}} \\\
(a) & ( b) & (c) \\
\end{tabular}
\caption{\it Panel (a): the system of coupled Landau-Stuart oscillators is stable if the eigenvalues of the Laplacian operator  
fall in the portion of the complex plane comprises in between the two solid lines. Symbols represent  the (randomly assigned) eigenvalues of the 
Laplacian that we aim at obtaining by means of  the procedure introduced in this paper. Panel (b): $\lambda_{Re}$ is plotted against $-\Lambda_{Re}=-\Re(\Lambda)$. This is an alternative way to show that the system is stable with the prescribed Laplacian spectrum. The solid line stands for the dispersion relation obtained in the continuum limit, when the discrete Laplacian is replaced by a standard differential operator.  Panel (c): the time evolution of the real components $\Re(W_j)$ is shown, with an appropriate color code.  Here, the weighted network which specifies the coupling between the nodes is obtained from the generated Laplacian, starting from the  assigned spectrum, with the procedure introduced in this paper.
Here, $c_1=3$, $c_2=2.4224$ and $K=1$. }
\label{fig:subfig1}
\end{figure*}

The system admits a homogeneous limit cycle solution in the form $W_{LC}(t)=e^{-ic_2t}$. To characterize the stability of the cycle, one can introduce a non homogeneous perturbation in polar coordinates as:
\begin{equation}
W_i(t)=W_{LC}[1+\rho_i(t)]e^{i\theta_i(t)}
\end{equation}
By linearizing around the limit cycle solution ($\rho_i(t)=0$, $\theta_i(t)=0$), one gets:  
\begin{equation}
\label{eqn:derivata_t}
\frac{d}{dt}
\begin{pmatrix}
\rho_j \\
\theta_j
\end{pmatrix}
=
\begin{pmatrix}
-2 & 0 \\
-2c_2 & 0
\end{pmatrix}
\begin{pmatrix}
\rho_j \\
\theta_j
\end{pmatrix}
+
\begin{pmatrix}
1 & -c_1 \\
c_1 & 1
\end{pmatrix}
\sum_k L_{jk}
\begin{pmatrix}
\rho_k \\
\theta_k
\end{pmatrix}
\end{equation}
To proceed further, expand the perturbations $\rho_j$ and $\theta_j$ on the Laplacian eigenvectors basis, that is
%\begin{multline}
%\sum_j\Delta_{ij}\phi_j^{(\alpha)}=\Lambda^{(\alpha)}\phi_i^{(\alpha)}, \quad \alpha=1,\dots,N
%\end{multline}
\begin{equation}
\begin{pmatrix}
\rho_j \\
\theta_j
\end{pmatrix}
=\sum_{\alpha=1}^{2N+1}
\begin{pmatrix}
\rho^{(\alpha)} \\
\theta^{(\alpha)}
\end{pmatrix}
e^{\lambda_{\alpha} t} (v_{(\alpha)})_i
\end{equation}
By inserting this expansion in \eqref{eqn:derivata_t} and using the relation
\begin{equation}
\sum_jL_{ij}(v_{(\alpha)})_j=\Lambda^{(\alpha)}(v_{(\alpha)})_i
\end{equation}
for $\alpha=1,\dots,2N+1$, we obtain a condition formally equivalent to the expression of the continuous dispersion relation
\begin{equation}
\label{eqn:disp_rel}
\lambda_{max}(\Lambda^{(\alpha)})=-\Lambda^{(\alpha)}-1+\sqrt{-c_1^2 \left({\Lambda^{(\alpha)}}\right)^2-2c_1c_2\Lambda^{(\alpha)}+1}
\end{equation}
If $\lambda_{Re}=\Re(\lambda_{max})$ is positive for some $\Lambda^{(\alpha)}$, the perturbation grows exponentially in time, and the initial homogeneous state proves unstable. 
Conversely, if $\lambda_{Re}=\Re(\lambda_{max}) < 0$, for every $\Lambda^{(\alpha)}$, the perturbation gets re-absorbed and the system converges back to the fully synchronized state. 
The condition $\lambda_{Re}<0$ can be further processed analytically, as discussed in \cite{cencetti3}. In particular, it can be shown that the latter condition is fulfilled, if the Laplacian eigenvalues fall in a specific portion of the parameter plane, which reflects the choice made for the reaction parameters $c_1$, $c_2$ and $K$. The region of interest is the one enclosed between the two solid lines, displayed in Fig. \ref{fig:subfig1}(a) for the specific selection of the parameters. The blue symbols depicted in Fig. \ref{fig:subfig1}(a) are randomly generated so as to fall in the  
region of the complex plane where stability holds. They represent the spectrum of the Laplacian that we seek to recover following the method illustrated above. In Fig. \ref{fig:subfig1}(b), $\lambda_{Re}$ is plotted against $-\Lambda_{Re}=-\Re(\Lambda)$, confirming the stability of the homogeneous solution. 

We now proceed by generating a Laplacian matrix, which is constructed so as to yield the spectrum depicted in Fig. \ref{fig:subfig1}(b). From this, we compute the corresponding adjacency matrix $A$ and use it to define the interactions between coupled oscillators, as follows from Eqs. (\ref{GL}). We then integrate numerically the governing equations, assuming the initial state to be a perturbation of the homogenous synchronized equilibrium. As expected, the perturbation fades away and the system regains its unperturbed, fully synchronized, equilibrium.  

\subsection{Coupled Kuramoto oscillators}

As a second example we set to study the Kuramoto model. Consider a system made of $2N+1$ oscillators, denote by  $\theta_i$ the phase of the $i$-th oscillator, and $\omega_i$ its natural frequency.
The oscillators evolve as dictated by the following system of $2N+1$ coupled differential equations:

\begin{equation}
\label{eqn:kuramoto}
\dot{\theta_i}=\omega_i+\sum_{j=1}^{2N+1}A_{ij}\sin (\theta_j-\theta_i) \qquad  i=1,.., 2N+1
\end{equation}

Here, $A_{ij}$ stands for the entries of the adjacency matrix $A$ which sets the interactions between pairs of oscillators. The matrix is, in principle, weighed, and may display positive and 
negative entries as reflecting the specific interaction (excitatory or inhibitory) being at play.

\begin{figure*}
\centering
\begin{tabular}{cc}
{\includegraphics[scale=0.4]{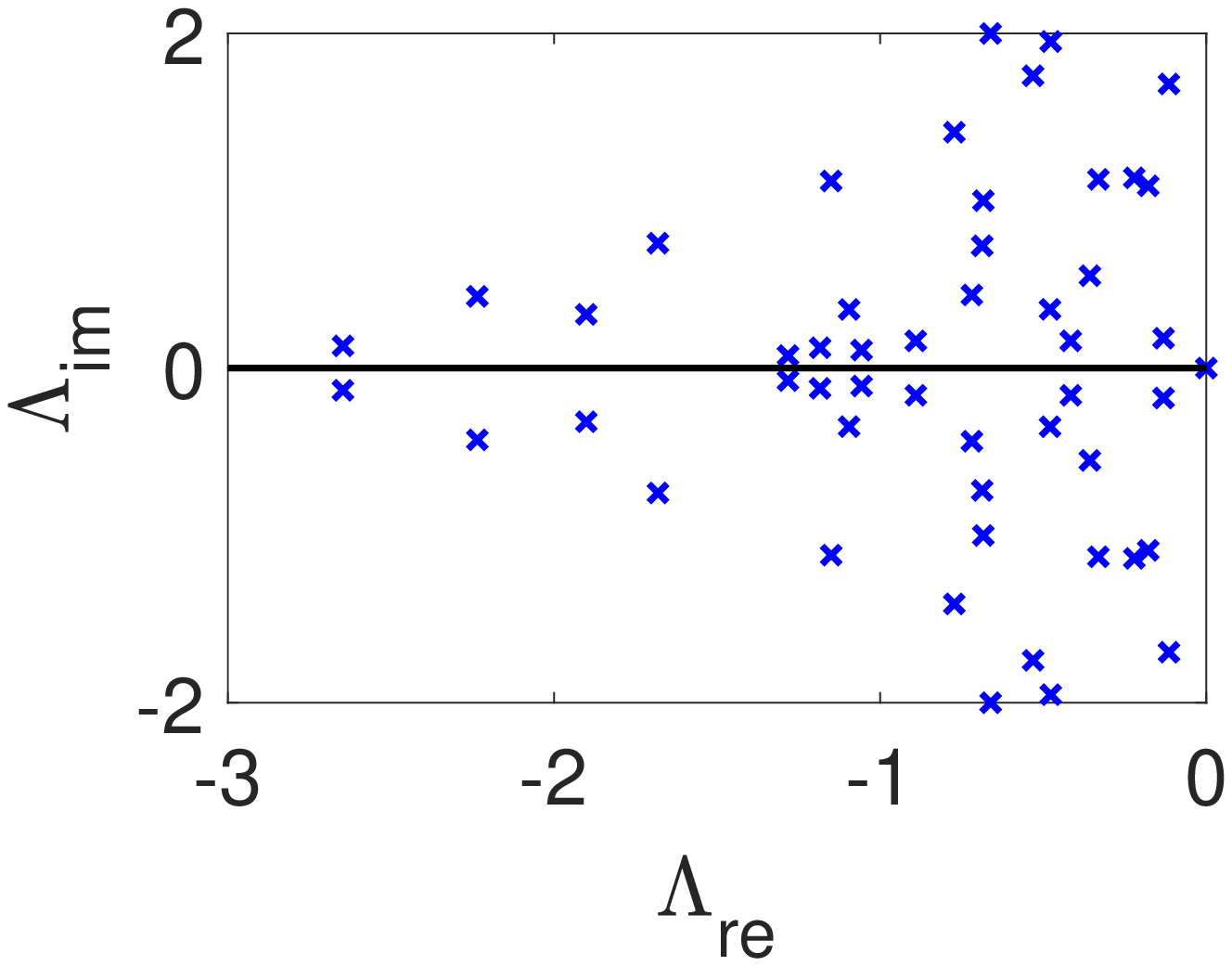}} &
{\includegraphics[scale=0.4]{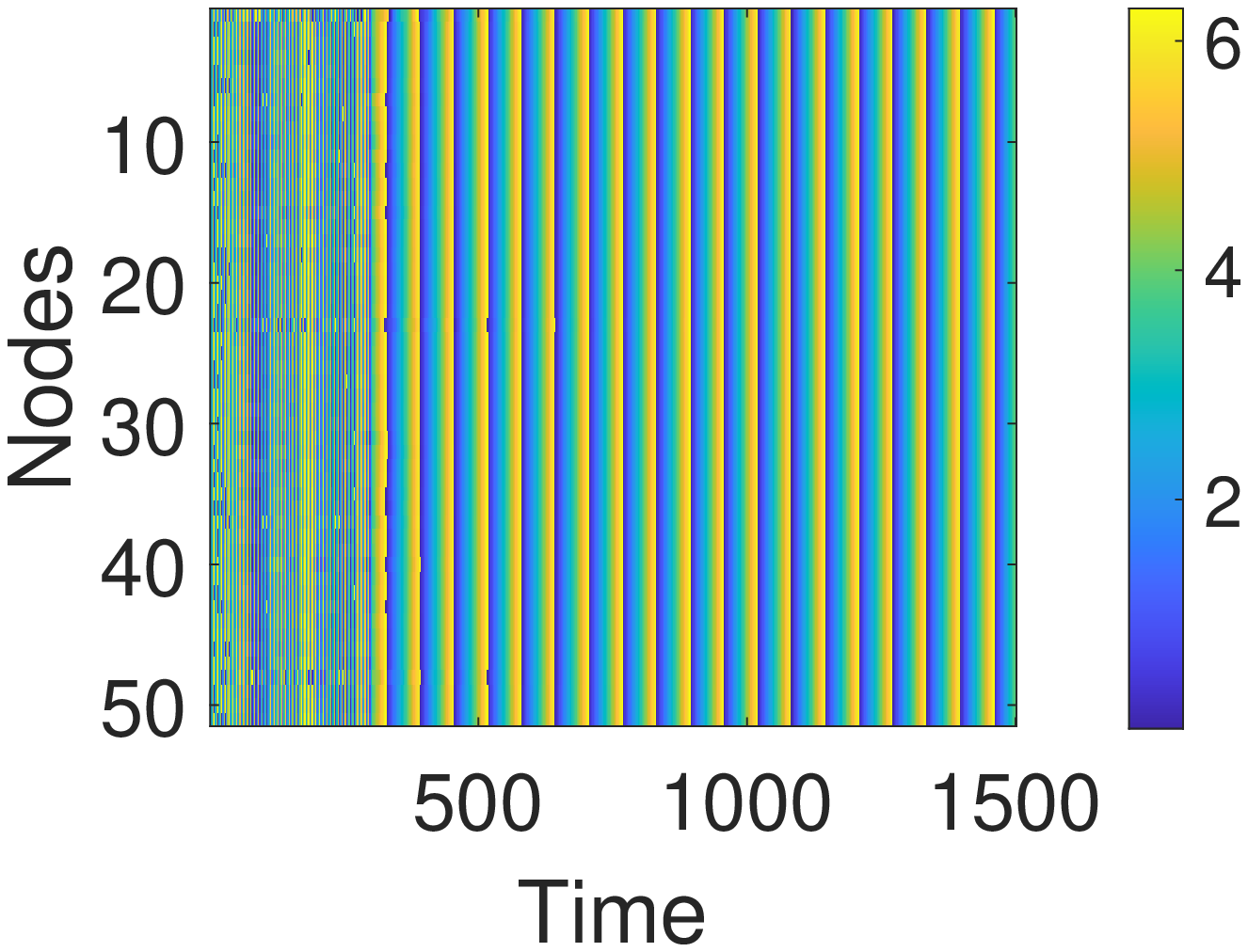}} \\
(a) & ( b)  \\
\end{tabular}
%\subfloat[][]
%{\includegraphics[scale=0.23]{stochastic_kuramoto.eps}}
\caption{\it Panel (a): the spectrum of the Laplacian operator that we seek to recover. The eigenvalues are distributed in the left portion of the complex plane to assure stability of the inspected Kuramoto model.  Panel (b): $\theta_i$ vs. $time$. Here, the adjacency matrix employed in the numerical integration follows from the determination of the Laplacian, via the procedure outlined in the main body of the paper and starting from the spectrum depicted in panel (a). As an initial condition, we perturb the homogenous solution (assumed $\Psi=0$) by a random heterogeneous amount. After a transient, the perturbation gets absorbed and the oscillators evolve in unison. Here $\Omega=2N+1=51$.}
\label{fig:subfig2}
\end{figure*}

As an additional assumption, we will here focus in the simplified setting where $\omega_i=\omega$ $\forall i$. We can then introduce the new variable $\psi_i=\theta_i-\omega t$, and write the governing equation in the equivalent form:
\begin{equation}
\label{eqn:kuramoto1} 
\dot{\psi_i}=\sum_{j=1}^{2N+1}A_{ij}\sin(\psi_j-\psi_i) \qquad  i=1,.., 2N+1
\end{equation}
A homogeneous solution always exists with $\psi_i=\Psi$ $\forall i$, and for any constant $\Psi \in [0, 2\pi)$, as it can be immediately checked by substitution. To assess the stability of 
the solution, one sets $\psi_i=\Psi+\delta_i$, and expands  \eqref{eqn:kuramoto1} at the leading order in the $\delta_i$. In this way, one gets:

\begin{equation}
\dot{\delta_i}=\sum_{j=1}^{2N+1}A_{ij}(\delta_j-\delta_i)=\sum_{j=1}^{2N+1}L_{ij}\delta_j
\end{equation}
where $L_{ij}=A_{ij}-k_i \delta_{ij}$ is the Laplacian operator which stems from $A_{ij}$. The stability of the simplified Kuramoto model here considered is  controlled by a linear system of the type introduced in (\ref{lin_syst}), with the obvious replacement of $x_i$ with $\delta_i$. The system proves hence stable if the (non trivial) eigenvalues of the Laplacian operator display negative real parts. Our aim, here,  is to generate a Laplacian (and therefore a matrix of binary weighted connections among oscillators) which warrants the stability of the system. To this end
we assign the eigenvalues (which appear in conjugate pairs) to belong to the negative portion of the complex plane, see Fig. \ref{fig:subfig2}(a). The null eigenvalue is clearly included into the spectrum. Running the procedure discussed in the first part of the paper, we obtain the corresponding Laplacian and compute the associated adjacency matrix. The Kuramoto model (\ref{eqn:kuramoto1}) is then integrated numerically by assuming the recovered expression for $A$. As predicted, the system is stable to external perturbations as one can clearly appreciate by inspection of Fig.  \ref{fig:subfig2}(b).

\section{Conclusions}

Studying the dynamics of an ensemble made of interacting units on a network is central for a large plethora of applications. In many cases of interests, individual units evolve under the 
influence of homologous constituents, the interaction stemming in general from binary exchanges. Distinct fundamental units are assigned to different nodes of the collection, paired via 
physical or virtual links. For a relevant subclass of problems, the stability of the ensuing equilibrium can be traced back to the spectrum of the Laplacian operator, computed from the adjacency matrix which defines the network arrangement. Symmetric networks yield Laplacian operators with real spectrum, while directionality in the couplings reflects in an imaginary spectrum. Methods exist which allow one to generate a symmetric network, hence Laplacian, with a prescribed real spectrum. Starting from these premises, we have proposed and tested a 
novel procedure to generate a (signed and directed) Laplacian which returns an a priori assigned (complex) spectrum. A special case has also been considered, which enables one to recover closed analytical expressions for the entries of the sought  Laplacian. Working in this setting, we can elaborate on the conditions that have to be matched for the ensuing Laplacian to solely display positive non diagonal elements. Dedicated sparsification procedures are also discussed to help removing unessential links in terms of their impact on the associated spectrum. The algorithm for Laplacian generation has been successfully tested with reference to two prototypical examples of  coupled oscillators.  Taken together, our work explores possible strategies for network generation with the emphasis placed on dynamical, rather than structural features. The dynamics is indirectly modulated by the spectrum of the Laplacian operator, which is here constraining the generation algorithm.

\appendix

\section{On the explicit expression of $S$}
\label{appS}

Given two arbitrary constants $\alpha$ and $\beta$, the following identity holds
\begin{equation}
(I+\alpha E)(I+\beta E)=I +(\alpha+\beta)E+\alpha\beta E^2
\end{equation}
for any matrix $E$. If $E$ is the matrix defined in \eqref{eqn:E} one gets:
\begin{equation}
E^2=NE
\end{equation}
and then
\begin{equation}
\label{eqn:identity}
(I+\alpha E)(I+\beta E)=I +(\alpha+\beta+\alpha\beta N)E
\end{equation}
If $\beta=-\frac{\alpha}{1+\alpha N}$, from \eqref{eqn:identity} we get
%\begin{equation}
%(I+\alpha E)(I+\beta E)=I 
%\end{equation}
%that is 
\begin{equation}
\label{eqn:inversion}
(I+\alpha E)^{-1}=(I+\beta E)
\end{equation}
The above result can be used to derive the structure of matrix $S$, as reported in the  main body of the paper. 
To this end, we begin by rewriting the matrices $A$ and $B$ as
\begin{equation}
A=[(1-i)E-iI]U=-i[I+(i+1)E]U 
\end{equation}
\begin{equation}
B=[(i+1)E+iI]U=i[I+(1-i)E]U
\end{equation}
From \eqref{eqn:inversion} we obtain
\begin{equation}
B^{-1}=-iU^{-1}[I+(i+1)E]^{-1}=-iU^{-1}\biggl[I+\frac{i-1}{1+(1-i)N}E\biggr]
\end{equation}
and then, after some calculations, we get
\begin{eqnarray*}
B+AB^{-1}A=i\biggl[I+(1-i)E+I+\frac{3i+2iN+1}{1+(1-i)N}E\biggr]U \\
=2i\biggl[I+\frac{1+i}{1+(1-i)N}E\biggr]U
\end{eqnarray*}
In conclusion,
\begin{equation*}
\label{eqn:S}
S=(B+AB^{-1}A)^{-1}=
-\frac{i}{2}U^{-1}\biggl[I-\frac{1+i}{1+2N}E\biggr]
\end{equation*}
because of 
\begin{equation}
\biggl[I+\frac{1+i}{1+(1-i)N}E\biggr]^{-1}=\biggl[I-\frac{1+i}{1+2N}E\biggr]
\end{equation}
which proves the results.

\section{About the computation of $L_{ij}$}
\label{app:contoL}
The aim of this section is to detail the computations needed to obtain explicitly the entries of the Laplace matrix $L$ under the assumption $U=qI$ starting thus from Eqs.~\eqref{eq:form1} and~\eqref{eq:form2}.

Let us begin by computing  the diagonal elements of $L$, namely  $L_{ii}$ for $i=1,..,N$, . For $i=1$, one gets:
\begin{equation}
\label{eqL11}
L_{11}=\sum_{k=2}^{N+1}V_{1k}D_{kk}W_{k1}
\end{equation}
From \eqref{eqn:v} and \eqref{eqn:w} we obtain for $k=2,\dots ,N+1$:
\begin{eqnarray}
V_{1k} &=&-1-i  \\
W_{k1}&=& \frac{-1+i}{2(2N+1)}  
\end{eqnarray}

Then
\begin{eqnarray}
L_{11}&=&2\sum_{k=2}^{N+1}\Re\biggl[(-1-i)D_{kk}\frac{-1+i}{2(2N+1)}\biggr]\\
&=& 2\sum_{k=2}^{N+1}\Re\biggl[D_{kk}(-1-i)\frac{-1+i}{2(2N+1)}\biggr]\\
&=&\frac{2}{2N+1}\sum_{k=2}^{N+1}\alpha_k
\end{eqnarray}
where use has been made of the identity $\Re(D_{kk})= \alpha_k$.

We can proceed in analogy for the others diagonal elements of the Laplacian matrix. In particular, for $s=2,\dots ,N+1$, we have that
\begin{equation}
W_{ss}=S_{ss}=\frac{-2Ni-1}{2(2N+1)}, 
\end{equation}
and then
\begin{eqnarray}
L_{ss}&=&2\sum_{k=2}^{N+1}\Re[D_{kk}V_{sk}W_{ks}]\\
&=&2\Re(D_{ss}V_{ss}W_{ss})\\
&=&2\Re\biggl[(\alpha_s+i\beta_s)i\frac{-2Ni-1}{2(2N+1)}\biggr]\\
&=&\frac{2N\alpha_s+\beta_s}{2N+1}
\end{eqnarray}
 Conversely, for $s=N+2, \dots ,2N+1$, we have
\begin{equation}
W_{ss}=iS_{ss}=i\frac{-2Ni-1}{2(2N+1)}, 
\end{equation}
and consequently:
\begin{eqnarray}
L_{ss}&=&2\Re\biggl[(\alpha_s-i\beta_s)i\frac{-2Ni-1}{2(2N+1)}\biggr]\\
&=&\frac{2N\alpha_s-\beta_s}{2N+1}
\end{eqnarray}
Let us proceed now with the other elements of the first row of $L$. For $t=2,\dots,N+1$ we can write
\begin{eqnarray}
L_{1t}&=&2\sum_{k=2}^{N+1}\Re(V_{1k}D_{kk}W_{kt})\\
&=&2\biggl[\Re(V_{1t}D_{tt}W_{tt})+\sum_{k=2,k~=t}^{N+1}\Re(V_{1k}D_{kk}W_{kt})\biggr]\\
&=&2\biggl\{\Re\bigg[(-1-i)(\alpha_t+i\beta_t)\frac{-2Ni-1}{2(2N+1)}\biggr]+\\
&&\sum_{k=2,k\neq t}^{N+1}\Re\bigg[(-1-i)(\alpha_k+i\beta_k)\frac{i-1}{2(2N+1)}\biggr]\biggr\}\\
&=&\frac{1-2N}{1+2N}\alpha_t-\beta_t+\frac{2}{2N+1}\sum_{k=2,k\neq t}^{N+1}\alpha_k
\end{eqnarray}
and for $t=N+2,\dots,2N+1$ we get
\begin{eqnarray}
L_{1t}&=&2\biggl\{\Re\bigg[(-1+i)(\alpha_t-i\beta_t)i\frac{-2Ni-1}{2(2N+1)}\biggr]+\\
&&\sum_{k=N+2,k\neq t}^{2N+1}\Re\bigg[(-1+i)(\alpha_k-i\beta_k)i\frac{i-1}{2(2N+1)}\biggr]\biggr\}\\
&=&\frac{1-2N}{1+2N}\alpha_t+\beta_t+\frac{2}{2N+1}\sum_{k=N+2,k\neq t}^{2N+1}\alpha_k\\
&=&\frac{1-2N}{1+2N}\alpha_t+\beta_t+\frac{2}{2N+1}\sum_{k=2,k\neq t-N}^{N+1}\alpha_k
\end{eqnarray}
For $s=2,\dots,N+1$ and $t=N+2,\dots,2N+1$ with $t\neq s-N$, we get:
\begin{eqnarray}
L_{st}&=&\sum_{k=2}^{2N+1}V_{sk}D_{kk}W_{kt}\\
&=&V_{ss}D_{ss}W_{st}+V_{s,s+N}D_{s+N,s+N}W_{s+N,t}\\
&=&i(\alpha_s+i\beta_s)(-i)\frac{-i-1}{2(2N+1)}+\\
&&(-i)(\alpha_s-i\beta_s)i\frac{i-1}{2(2N+1)}\\
&=&\frac{-\alpha_s+\beta_s}{2N+1}
\end{eqnarray}
while, for $s=N+2,\dots ,2N+1$ and $t=2,\dots,N+2$ with $t\neq s-N$, the following expression holds:
\begin{eqnarray}
L_{st}&=&V_{ss}D_{ss}W_{st}+V_{s,s-N}D_{s-N,s-N}W_{s-N,t}\\
&=&(\alpha_s-i\beta_s)\frac{-i-1}{2(2N+1)}+(\alpha_s+i\beta_s)\frac{i-1}{2(2N+1)}\\
&=&-\frac{\alpha_s+\beta_s}{2N+1}
\end{eqnarray}
For $s=2,\dots,N+1$
\begin{eqnarray}
L_{s,s+N}&=&\\
&& V_{ss}D_{ss}W_{s,s+N}+V_{s,s+N}D_{s+N,s+N}W_{s+N,s+N}\\
&=&i(\alpha_s+i\beta_s)(-i)\frac{2Ni-1}{2(2N+1)}+\\
&&(-i)(\alpha_s-i\beta_s)i\frac{-2Ni-1}{2(2N+1)}=\\
&&-\frac{\alpha_s+2N\beta_s}{2N+1}
\end{eqnarray}
while, for $s=N+2,\dots,2N+1$, we obtain:
\begin{eqnarray}
L_{s,s-N}&=&\\
&&V_{ss}D_{ss}W_{s,s+N}+V_{s,s-N}D_{s-N,s-N}W_{s-N,s-N}\\
&=&(\alpha_s-i\beta_s)\frac{2Ni-1}{2(2N+1)}+(\alpha_s+i\beta_s)i\frac{-2Ni-1}{2(2N+1)}=\\
&&-\frac{\alpha_s-2N\beta_s}{2N+1}
\end{eqnarray}
For $t=2,\dots,N+1$
\begin{eqnarray}
L_{t1}&=&2\Re(V_{tt}D_{tt}W_{t1})=\\
&&2\Re\biggl[i(\alpha_t+i\beta_t)\frac{i-1}{2(2N+1)}\biggr]=\\
&&-\frac{\alpha_t-\beta_t}{2N+1}
\end{eqnarray}
and for $t=N+2,\dots,2N+1$
\begin{eqnarray}
L_{t1}&=&2\Re(V_{tt}D_{tt}W_{t1})\\
&=&2\Re\biggl[(\alpha_t-i\beta_t)\frac{-i-1}{2(2N+1)}\biggr]\\
&=&\frac{-\alpha_t-\beta_t}{2N+1}
\end{eqnarray}
For $t,s=2,\dots,N+1$ and $s\neq t$
\begin{eqnarray}
L_{st}&=&2\Re(V_{ss}D_{ss}W_{st})\\
&=&2\Re\biggl[i(\alpha_s+i\beta_s)\frac{i-1}{2(2N+1)}\biggr]\\
&=&\frac{-\alpha_s+\beta_s}{2N+1}
\end{eqnarray}
and, finally, for $s,t=N+2,\dots,2N+1$ and $s\neq t$, one gets:
\begin{eqnarray}
L_{st}&=&2\Re(V_{ss}D_{ss}W_{st})\\
&=&2\Re\biggl[(\alpha_s-i\beta_s)i\frac{i-1}{2(2N+1)}\biggr]=\\
-\frac{\alpha_s+\beta_s}{2N+1}
\end{eqnarray}

Summing up, we have here provided closed-form analytical expressions for all the entries of the Laplacian matrix, as a function of the assigned spectrum.  

\section{Positiveness of $L$}
\label{app:positL}

The aim of this section is to work out the algebraic steps needed to rewrite~\eqref{eqn:sistema} in the simpler form given by (\ref{eqn:condition_alpha1}) and (\ref{eqn:condition_beta1}).

Let us thus rewrite~\eqref{eqn:sistema}
\begin{equation*}
\begin{cases}
2N\alpha_t<\beta_t<-2N\alpha_t  \\
\beta_t>-\frac{1-2N}{2N+1}\alpha_t-\frac{2}{2N+1}\sum_{k=2,k\neq t}^{N+1}\alpha_k \\
\beta_t<\frac{1-2N}{2N+1}\alpha_t+\frac{2}{2N+1}\sum_{k=2,k\neq t}^{N+1}\alpha_k  \\
\alpha_t<\beta_t<-\alpha_t  \\
\frac{\alpha_t}{2N}<\beta_t<-\frac{\alpha_t}{2N}  \\
\end{cases}
\end{equation*}
where the inequalities hold for $t=2,\dots,N+1$. Then the second and the third conditions of system (\ref{eqn:sistema}) can be matched 
 simultaneously provided that:
\begin{equation}
\alpha_t<\frac{2}{2N-1}\sum_{k\neq t}\alpha_k \quad \forall t=2,\dots,N+1
\end{equation}

Notice that: 
\begin{equation}
2N\alpha_t<\alpha_t<\frac{\alpha_t}{2N}
\end{equation}
holds for arbitrary values of $\alpha_t<0$ and $N$. Hence,  system \eqref{eqn:sistema} simplifies as follows:
\begin{equation}
\label{eqn:sistema2}
\begin{cases}
\frac{\alpha_t}{2N}<\beta_t<-\frac{\alpha}{2N}\\
\alpha_t<\frac{2}{2N-1}\sum_{k\neq t}\alpha_k \\
\beta_t>-\frac{1-2N}{2N+1}\alpha_t-\frac{2}{2N+1}\sum_{k=2,k\neq t}^{N+1}\alpha_k\\
\beta_t<\frac{1-2N}{2N+1}\alpha_t+\frac{2}{2N+1}\sum_{k=2,k\neq t}^{N+1}\alpha_k 
\end{cases}
\end{equation}
for $t=2,\dots,N+1$.

Focus now on the conditions for $\beta_t$. We assume that the following condition holds:
\begin{equation}
\label{eqn:ansatz}
\frac{1-2N}{2N+1}\alpha_t+\frac{2}{2N+1}\sum_{k=2,k\neq t}^{N+1}\alpha_k>-\frac{\alpha_t}{2N}
\end{equation}
and we set to explore its consequences. Eq. (\ref{eqn:ansatz}) yields:
\begin{equation}
\label{eqn:condition_alpha}
\alpha_t<\frac{4N}{4N^2-4N-1}\sum_{k\neq t}\alpha_k=\frac{1}{N-1-\frac{1}{4N}}\sum_{k\neq t}\alpha_k
\end{equation} 

For $N>1$,  conditions \eqref{eqn:sistema2} maps therefore in the following equivalent system:
\begin{equation}
\label{eqn:sistema3}
\begin{cases}
\frac{\alpha_t}{2N}<\beta_t<-\frac{\alpha_t}{2N}\\
\alpha_t<\frac{2}{2N-1}\sum_{k=2,k\neq t}^{N+1}\alpha_k\\
\alpha_t<\frac{1}{N-1-\frac{1}{4N}}\sum_{k\neq t}\alpha_k
\end{cases}
\end{equation}
Remark that:
\begin{equation}
\frac{1}{N-1-\frac{1}{4N}}\sum_{k=2,k\neq t}^{N+1}\alpha_k<\frac{2}{2N-1}\sum_{k=2,k\neq t}^{N+1}\alpha_k
\end{equation}
due to the inequality
\begin{equation}
\frac{1}{N-1-\frac{1}{4N}}>\frac{2}{2N-1}
\end{equation}
which holds for each  $N>1$. Hence,  system \eqref{eqn:sistema3} takes the form:
\begin{equation}
\label{eqn:prime}
\begin{cases}
\frac{\alpha_t}{2N}<\beta_t<-\frac{\alpha_t}{2N}\\
\alpha_t<\frac{1}{N-1-\frac{1}{4N}}\sum_{k\neq t}\alpha_k
\end{cases}
\end{equation}

Then, \eqref{eqn:prime} has no solutions, under the working hypothesis that we have put forward to deriving it. In fact: 
\begin{equation}
\label{eqn:calculus1}
\alpha_t+\frac{1}{N-1-\frac{1}{4N}}\alpha_t<\frac{1}{N-1-\frac{1}{4N}}\sum_k\alpha_k
\end{equation}
that is
\begin{equation}
\label{eqn:calculus2}
\alpha_t<\frac{1}{N-\frac{1}{4N}}\sum_k\alpha_k=\frac{4N}{4N^2-1}\sum_k\alpha_k
\end{equation}
and summing on every $t$ we get
\begin{equation}
\sum_t\alpha_t<\frac{4N^2}{4N^2-1}\sum_k\alpha_k
\end{equation}
which, in turn, implies $\frac{4N^2}{4N^2-1}<1$, a condition that is obviously never met. We now go back to revise ansatz \eqref{eqn:ansatz}, and consider the alternative scenario:
\begin{equation}
\frac{1-2N}{2N+1}\alpha_t+\frac{2}{2N+1}\sum_{k=2,k\neq t}^{N+1}\alpha_k<-\frac{\alpha_t}{2N}
\end{equation}
Then, \eqref{eqn:sistema2} becomes
\begin{equation}
\label{eq:second}
\begin{cases}
\beta_t>-\frac{1-2N}{2N+1}\alpha_t-\frac{2}{2N+1}\sum_{k=2,k\neq t}^{N+1}\alpha_k\\
\beta_t<\frac{1-2N}{2N+1}\alpha_t+\frac{2}{2N+1}\sum_{k=2,k\neq t}^{N+1}\alpha_k \\
\alpha_t<\frac{2}{2N-1}\sum_{k=2,k\neq t}^{N+1}\alpha_k\\
\alpha_t>\frac{1}{N-1-\frac{1}{4N}}\sum_{k\neq t}\alpha_k
\end{cases}
\end{equation}
Following a path analogous to the one discussed above, we get:
\begin{equation}
\label{eqn:condition_alpha2}
\frac{4N}{4N^2-1}\sum_k\alpha_k<\alpha_t<\frac{2}{2N+1}\sum_k\alpha_k
\end{equation}
and
\begin{equation}
\label{eqn:condition_beta2}
\alpha_t-\frac{2}{2N+1}\sum_k\alpha_k<\beta_t<\\
-\alpha_t+\frac{2}{2N+1}\sum_k\alpha_k
\end{equation}

\end{document}